\documentclass{article}%
\usepackage[utf8]{inputenc}%
\usepackage{csquotes}%
\usepackage{authblk}
\usepackage{url}%
\usepackage{geometry}%
\usepackage{booktabs}%
\usepackage{color}%
\usepackage{soul}%
\usepackage[normalem]{ulem}%
\usepackage{comment}%
\usepackage{setspace}%
\usepackage{amsmath}
\usepackage[hidelinks]{hyperref}%
\usepackage[export]{adjustbox}%
\usepackage[inline]{enumitem}%
\usepackage{rotating}%
\usepackage{tabularx}%
\usepackage{framed}%
\usepackage{makecell}%
\usepackage{color}%
\usepackage{caption}%
\usepackage{subcaption}%
\usepackage{footmisc}

\usepackage[natbibapa]{apacite}
\usepackage[hidelinks]{hyperref} 
\hypersetup{
    colorlinks=false,
    linkcolor=black,
    citecolor=black,
    urlcolor=black,
}
\input{prompts.tex}%
\renewcommand{\hl}[1]{#1}
\begin{document}%
\title{Prompting AI Art: An Investigation into the Creative Skill of Prompt Engineering}%
%
%


\author[1]{Jonas Oppenlaender} 
\author[2]{Rhema Linder}
\author[3]{Johanna Silvennoinen}
\affil[1]{Independent Researcher, \texttt{oppenlaenderj@acm.org}}
\affil[2]{University of Tennessee, Knoxville, \texttt{rlinder@utk.edu}}
\affil[3]{University of Jyv\"askyl\"a, \texttt{johanna.silvennoinen@jyu.fi}}

\date{}

\maketitle%

\begin{abstract}%
\noindent
We are witnessing a novel era of creativity where anyone can create digital content via prompt-based learning
(known as prompt engineering).
This paper investigates prompt engineering as a novel creative skill for 
creating AI art with text-to-image generation.
In three consecutive studies, we explore whether crowdsourced participants 
can 1) discern prompt quality, 2) write prompts, and 3) refine prompts.
We find that participants could evaluate prompt quality and crafted descriptive prompts, but they lacked style-specific vocabulary necessary for effective prompting.
This is in line with our hypothesis that prompt engineering is a new type of skill that is non-intuitive and must first be acquired (e.g., through means of practice and learning) before it can be used.
Our studies deepen our understanding of prompt engineering and chart future research directions.
We conclude by envisioning four potential futures for prompt engineering.
\end{abstract}%

\section{Introduction}%
%
We are entering an era in which anybody can generate digital images from text --- a democratization of art and creative production.
In this novel creative era, humans work 
within a human-computer co-creative framework \citep{062-iccc20.pdf}.
Emerging digital technologies will co-evolve with humans in this digital revolution, which requires the renewal of human capabilities and competences \citep{GenZ,10447318.2023.2171489} and a human-centered design process \citep{10447318.2024.2345430}.
One increasingly important human skill in this context is \textit{prompting} due to it providing an intuitive language-based interface to artificial intelligence (AI).
Prompting (or ``prompt engineering'') is the skill and practice of writing inputs (``prompts'') for generative models \citep{guidelines,aiartcreativity}.
Prompt engineering is iterative and interactive~--- a dialogue between humans and AI in an act of co-creation.
As generative models become more widespread, prompt engineering has become an important research area on how humans interact with AI \citep{prompt-programming,2209.01390.pdf,2212.07476.pdf,2022.acl-demo.9.pdf,3491101.3503564.pdf,PERCEPTIONS,guidelines,2209.11486.pdf,3544548.3580969.pdf,10447318.2024.2311971}.

One area where prompt engineering has been particularly useful is the field of digital visual art. Image generation is often utilised in the later stage of the creation process in generating solutions, not in the ideation phase \citep{lee2024and}. State-of-the-art image generation systems, such as OpenAI's DALL-E~\citep{DALLE2}, Midjourney~\citep{midjourney}, and
 Stable Diffusion~\citep{SD},
have been trained on large collections of text and images collected from the World Wide Web. These systems can synthesize high-quality images in a wide range of artistic styles from textual input prompts \citep{guidelines,artiststudies,modifiers,aiartcreativity}.
Practitioners of text-to-image generation often use prompt engineering to improve the quality of their digital artworks \citep{guidelines}. Within the community of practitioners, certain keywords and phrases have been identified that act as ``prompt modifiers'' \citep{modifiers}. These keywords can, if included in a prompt, improve the quality of the generative model's output or make images appear in a specific artistic style \citep{guidelines,artiststudies,aiartcreativity}.
While a short prompt may already produce impressive results with these generative systems, the use of prompt modifiers can help practitioners unlock the systems' full potential \citep{modifiers,aiartcreativity,2303.04587.pdf}. The skillful application of prompt modifiers can distinguish expert practitioners of text-to-image generation from novices.



However, how people interact with image-generation systems is a relatively unexplored phenomenon \citep{kim2024journey}.  In addition, whether prompt engineering is an intuitive skill or whether this skill must be acquired (e.g., through means of practice and iterative learning) 
has, so far, not been investigated.
Investigating the skill of prompt engineering is important for several reasons.

For the field of AI art it is important to inform the development of future image generation systems and interfaces.
A look at Stable Diffusion's Discord channel\footnote{https://discord.com/channels/1002292111942635562/}
provides 
anecdotal
evidence that some prompts and keywords combinations 
circulating in the community of practitioners are not intuitive.
    Such keywords include, for instance, the modifier \textit{``by Greg Rutkowski''} or other popular keywords, such as 
    \textit{``smooth,''} \textit{``elegant,''} \textit{``luxury,''} \textit{``octane render,''} ``\textit{trending on},'' and \textit{``artstation''}.
    These keywords are often used in combination with each other to boost the quality of generated images~\citep{modifiers}, resulting in unintuitive keyword combinations that a human user would likely never have chosen to describe the image to another human.
With these many keywords comes a loss of control over the outcome. There is a high randomness to the outcome of text-to-image generation, and controlling the image generation (without resorting to additional tools, such as ControlNet \citep{ControlNet}, is difficult, even for experts in prompt engineering).
    Keywords and modifiers are commonly 
    applied by practitioners in the AI art community, but may confront laypeople with challenges of understanding the effect of modifiers on the resulting image.
Further confounding the problem is that keywords in prompts can affect both the subject and style of a generated image simultaneously.

For the field of education, investigating the skill of prompt engineering is important, since the popularity and importance of the practice of prompt engineering is growing. Prompt engineering has been proven effective in solving difficult problems that were previously too complex to solve. This property makes prompt engineering useful in industry contexts. Therefore, a legitimate question to ask if whether prompt engineering should be included as a subject in school and higher education curricula. To inform this decision, one needs to know more about prompt engineering as a skill -- what is its learning curve, how many hours would it take to learn, or is it even a skill?
If people can just simply apply this language-based skill without much learning is important information to have in this context.

Furthermore, whether prompt engineering is a skill that humans apply intuitively or whether it is a new type of skill that needs to be acquired is important not only for the field of AI art and education, but also for research on human-AI interaction and the future of work in general.
Consider, for instance, the sector of (higher) education and its future curricula. Agents based on language models (LMs) based have become vastly popular and promise to solve complex problems that previously required software engineering expertise \citep{wu2024autogen}. Such LM-based agents heavily rely on prompt engineering. The question, then, is whether current educational curricula should be extended with teaching on prompt engineering.
However, if prompt engineering was a skill that everyone with sufficient knowledge of the English language could just apply without having to learn this skill, it would mean that extending the curricula with prompt engineering would not be necessary. An extension would take away from the time to teach other, more important, skills. On the other hand, if there is a learning curve to it, or if the skill proves to be transferable enough to be valuable, then its adoption in curricula would be warranted.

A source for unintuitiveness of the current practice of prompt engineering is a potential misalignment between the human-written prompts and the way in which text-to-image models interpret prompts.
    Compared to how we humans understand a prompt and its constituents, text-to-image generative models may attach very different meanings to some keywords in the prompt.
Further, many AI-generated images are shared on social media, often with stunning results.
However, if what we see on social media is the result of the application of prompt engineering by skilled experts, then the generative content that we encounter on social media could be skewed by a small group of highly skilled practitioners.
Or from another perspective, if prompt engineering is an acquired skill that requires expertise and training, this could give rise to novel creative professions with implications for the future of work.
On the other hand, we run the risk of assigning too much importance to prompting as a method for interacting with generative models if prompt engineering is an innate ability and an intuitive artistic skill that is acquired quickly~\citep{GenAI,PERCEPTIONS}.

In this paper, we explore the creative skill of prompt engineering in three studies with a total of 227~participants recruited from Amazon Mechanical Turk (MTurk), a 
crowdsourcing platform.
The timing of our three studies is significant (see Table \ref{tab:studiesoverview}. The studies were conducted at a time when text-to-image generation was still relatively unknown. This allowed us to study whether participants could apply prompt engineering intuitively, or whether prompt engineering is a skill that must be learned through many iterations. We expand on the timing of this study in Section \ref{sec:futurework}.

%
\begin{table*}[htb]%
\caption{Overview of the \hl{three} studies presented in this paper.}%
\label{tab:studiesoverview}%
\small%
\begin{tabularx}{\textwidth}{
    c
    c
    c
    X 
    X
}%
\toprule%
    \makecell{\textbf{No.} }
    &
    \makecell{\textbf{Participants}}
    &
    \makecell{\hl{\textbf{Study} \\ \textbf{date}}}
    &
    \makecell[l]{\textbf{Study purpose}}
    &
    \makecell[l]{\textbf{Research question}}
\\%
\midrule%
	     1
	 &
         52
	 &
  \hl{May 19--23, 2022}
     & 
        Test participants' understanding of prompt quality 
     &
         Are participants able to tell the quality of an image from the textual input prompt?
\\
    	 2
	 &
         125
	 &
  \hl{June 12--23, 2022}
     &
         Test participants' ability to write prompts
     &
         Can participants effectively write prompts to create digital artworks?
\\
    	 3
	 &
         50
	 &
  \hl{June 20--27, 2022}
     &
         Test participants' ability to revise their own prompts
     &
         Can participants improve their prompts to generate better digital artworks?
\\%
\bottomrule%
\end{tabularx}%
\end{table*}%
%




In \textbf{Study~1}, we explore participants' understanding of how a text-to-image generation system produces images of varying quality depending on the phrasing of input prompts.
A feeling of what contributes to the quality of a prompt could enable participants to write prompts and create high-quality images.
In our within-subject experiment, participants separately rated the aesthetic appeal of textual prompts and matching images generated with a text-to-image generation system.
We hypothesize that a high degree of consistency within the participants' two ratings may point toward there being a strong understanding of what makes a ``good'' prompt.

\textbf{Key insights from Study 1:}
We find participants are able to grasp what makes a ``good'' prompt. Being able to discern good from bad prompts would, in theory, allow participants to write effective prompts.

In \textbf{Study~2}, we test the insights from the above two studies in practice. We invite participants to apply their knowledge and expertise by writing three input prompts for a text-to-image generation system with the specific aim of creating a digital artwork (and without seeing the generated images).
We analyze participants' use of descriptive language and the use of prompt modifiers that could influence the quality and style of the resulting artworks.
In \textbf{Study~3}, we then invite the same participants who participated in the Study~2 to review the images generated from their own prompts. Each participant was asked to improve their prompts with the specific task of creating an artwork of high visual quality.
Our hypothesis is that if prompt engineering is an intuitive skill innate to humans, participants will be able to apply it immediately. On the other hand, if participants are not able to significantly improve their images due to few interactions with the text-to-image generation system within our studies, then this may indicate that the skill of prompt engineering needs to be acquired before it can be applied in practice.

\textbf{Key insights from studies 2 and 3:}
We find that while participants were able to describe artworks in rich descriptive language, almost none of the participants used specific keywords to adapt the style of their artworks or modify the images in other ways.
Moreover, participants were not able to significantly improve the quality of the artworks in the follow-up study.
This points to prompt engineering being a non-intuitive skill that people first need to acquire before it can be applied in meaningful ways.

In summary, our three studies find that while laypeople participants had the prerequisites to write prompts for AI art and were good at crafting descriptive prompts, they lacked style-specific vocabulary necessary for effective prompt engineering.
We conclude by speculating on four potential futures for prompt engineering. 

\section{Related Work}%
\label{sec:related-work}%
%
%
\subsection{Text-to-Image Generation with Deep Learning}%
\label{sec:background}%
%
Text-to-image generation is a type of generative deep learning technology that allows users to create images from text descriptions. This technology has gained significant interest since early 2021, when OpenAI published the results of DALL-E \citep{DALL-E} and the weights of their CLIP model \citep{CLIP}. CLIP is a multi-modal model trained on over 400 million text and image pairs from the Web. The model can be used in text-to-image generation systems to guide the generation of high-fidelity images.
Many approaches and architectures for image generation with deep learning have since been developed, such as diffusion models \citep{2105.05233.pdf}. These approaches typically use machine learning models trained with contrastive language-image techniques using training data scraped from the Web. These systems are text-conditional, meaning they use text as input for image synthesis. This input, known as ``prompt,'' describes the image to the system, which then generates one or more images without further input.

\subsection{Prompt Engineering for AI Art}%
\label{sec:promptengineering}%
The practice of crafting input prompts is referred to as prompt engineering (or prompting for short).
In this section, we explain the `engineering' character of prompt engineering and how prompt engineering is applied for generating AI art.%
%
%
\subsubsection{The engineering character of prompting}
The term prompt engineering was originally coined by 
Gwern Branwen in the context of writing textual inputs for OpenAI's GPT-3 language model~\citep{guidelines}.
`Engineering,' in this case, does not refer to a hard science as found in science, technology, engineering, and mathematics (STEM) disciplines.
    Prompt engineering is a term that originates from within the online community of practitioners.
    Practitioners include artists and creative professionals, but also novices, amateurs, and more serious ``Pro-Ams''~\citep{2556288.2557298.pdf} aiming, for instance, to sell their creations as digital art based on non-fungible tokens (NFTs) \citep{nft}.
    Not every member of this online community may identify as a prompt engineer. An alternative self-understanding could be 
    ``promptist'' \citep{promptism} or ``AI artist'' \citep{Zylinska_2020_AI-Art.pdf}.
One aspect of prompt engineering that relates to its engineering character is that it often involves systematic experimentation through trial and error~\citep{guidelines}. The challenge for the prompt engineer is not only to find the right terms to describe an intended output and the right m, but also to anticipate how other people would have described and reacted to the output on the World Wide Web.


\subsubsection{Prompt engineering for creating AI art}
\label{sec:promptmodifiers}%
Art generated by artificial intelligence, or ``AI art'' \citep{Zylinska_2020_AI-Art.pdf},
has become a popular application for prompt engineering \citep{aiartcreativity}.
An online community around AI art has formed, sharing images and prompts on various platforms. Within this community, certain practices for writing prompts have emerged. For example, prompts often follow a specific pattern, such as the following template \citep{travelersguide}:
\begin{quote}
  \textit{[Medium] [Subject] [Artist(s)] [Details] [Image repository support]}
\end{quote}
A typical prompt could be \citep{zippy}:
\begin{quote}%
    \textit{A beautiful painting of a singular lighthouse, shining its light across a tumultuous sea of blood \underline{by greg rutkowski} \underline{and thomas kinkade}, \underline{trending on artstation}.}%
\end{quote}%

Prompt modifiers, such as the underlined terms above, are added to a prompt to influence the resulting image in a specific way \citep{guidelines,modifiers,aiartcreativity}. Prompt modifiers are an important technique in prompt engineering for AI art because they allow the prompt engineer to control the output of the text-to-image generation system. Prompt modifiers may make the resulting images subjectively more aesthetic and attractive~\citep{guidelines,aiartcreativity}.

Different types of prompt modifiers are used in the AI art community \citep{modifiers}, but the two most common types of modifiers affect the style and quality of images. These prompt modifiers consist of specific keywords and phrases that have been found to modify the style or quality of an image (or both).
Modifiers that affect the quality of images can be referred to as `quality boosters' \citep{modifiers}, and include phrases such as \textit{``trending on artstation,'' ``unreal engine,'' ``CGSociety,'' ``8k,'' and ``postprocessing.''}
Style modifiers affect the style of an image and can include a wide variety of open domain keywords and phrases, such as \textit{``oil painting,'' 
``in the style of surrealism,''  or ``by James Gurney''}~\citep{modifiers}.

Human-centered research on prompt engineering for text-to-image synthesis in the field of Human-Computer Interaction (HCI) is still in its early stages. 
The study by \citet{guidelines} 
on subject and style keywords in textual input prompts mentioned that without knowledge of prompt modifiers, users must engage in ``brute-force trial and error''. The authors presented design guidelines to help people produce better results with text-to-image generative models \citep{guidelines}.
\citet{3527927.3532792.pdf} 
conducted an experiment on using images as visual input prompts, resulting in design guidelines for improving subject representations in AI art. Besides these guidelines, there are also many community-created resources that offer guidance for novices and practitioners of AI art, such as the ``Traveler's Guide to the Latent Space''~by \citet{travelersguide}, Zippy's ``Disco Diffusion Cheatsheet''~\citep{zippy}, and 
the ``Disco Diffusion Artist Studies'' by \citet{artiststudies}.
These resources provide a wealth of information about prompt modifiers for producing high-quality visual artifacts.


\subsection{Prompt Engineering as a Skill}%
\label{sec:skill}%

Merriam-Webster defines `skill' as ``the ability to use one's knowledge effectively and readily in execution or performance'' and ``a learned power of doing something competently: a developed aptitude or ability'' \citep{skill}.
This definition captures the essence of skill as not only having knowledge, but also the capability to apply the knowledge effectively in practical situations and in real-world contexts.

In our work, we define `skill in prompt engineering' as the ability to effectively utilize language and 
prior knowledge to craft prompts that 
guide generative models towards desired outputs. This encompasses not only the basic knowledge of the relevant language syntax but also the strategic use of prompt modifiers --- elements that refine or alter the direction of generative outputs. Our study aims to investigate whether individuals, particularly participants recruited from a crowdsourcing platform, possess the necessary foundational knowledge to write effective prompts. More importantly, we assess if these participants can translate this knowledge into practical application, demonstrating a skilled use of prompt modifiers in the context of text-to-image generation for AI art. The inability to effectively apply knowledge in practice may suggest a lack of skill, underscoring the need for knowledge acquisition and refinement through learning and training. This approach aligns with our objective to elucidate 
whether prompt engineering is an innate ability or whether it must be acquired (e.g., through practice and learning).


\subsection{Prior Work on Applying Skill in Practice}%

Our research is related to prior work focusing on how individuals acquire 
skills and how they apply them in practice. In particular, learning how to use web search is a related area of work. This section is structured around three main themes identified in the literature: the divergence between search and domain expertise, the strategies involved in how people rewrite a query after a failed attempt, and teaching how to improve query authorship.

\subsubsection*{Divergence Between Search and Domain Expertise}
The first theme addresses the relationship between domain expertise and search expertise. \citet{EXT-White} provide an insightful analysis of how domain knowledge influences web search behavior. Their work demonstrates that domain experts and novices exhibit markedly different search strategies and outcomes.
This finding is supported and extended by \citet{EXT-Wood}, who explore the relative contributions of domain knowledge and search expertise in conducting effective internet searches \citep{EXT-Huang}. These studies underscore the distinct nature of search expertise, separate from domain-specific knowledge, highlighting its importance in effective information retrieval.

\subsubsection*{Query Reformulation Strategies}
The second theme revolves around how individuals modify their search queries following unsuccessful attempts. \citet{EXT-Huang} offer a comprehensive examination of query reformulation strategies in web search logs. Their analysis reveals the common patterns and tactics users employ when their initial search queries fail to yield desired results. This research is crucial in understanding the adaptive behaviors of users in response to the challenges they encounter during web searches.

\subsubsection*{Educational Approaches to Query Authorship}
The third theme relates to methods for teaching effective query formulation. \citet{EXT-Bateman} contribute significantly to this area through their development of the Search Dashboard tool. Their study examines how tools that facilitate reflection and comparison can impact users' search behaviors, leading to more effective search strategies. This work is particularly relevant for designing educational interventions and tools aimed at enhancing the search skills of users.

Our studies draw upon and contribute to these existing bodies of work. We extend the understanding of how individuals apply their knowledge in writing prompts, and how they reformulate a prompt in an attempt to improve upon a first prompt.
Both are crucial aspects of information literacy in the digital age.
\section{Study 1: Understanding Prompt Engineering}%
\label{sec:study2}%
We conducted a within-subject experiment to study participants' understanding of prompt engineering. Participants were asked to rate the textual prompts and the corresponding AI-generated images. We hypothesize that participants with a strong understanding of prompt engineering would exhibit a high consistency between the ratings in the two modalities.
In other words, if someone can predict the aesthetic appeal of an image from its textual prompt, they likely have a good sense of how prompt engineering works.
The study design reflects the knowledge that prompt engineers would draw on in practice: first write a prompt with the intention to produce a high quality image, then observe and assess the quality of the resulting image. A good understanding of textual prompts is crucial for predicting how well a prompt will perform.
For instance, prompts incorporating rich descriptive language and multiple prompt modifiers tend to yield artworks of higher quality compared to prompts lacking these attributes.\footnote{Note that some state-of-the-art image generation systems, like Midjourney, are ``greedy'' and will try to turn any input into an aesthetic artwork, even if the prompt is short or non-descriptive. See Section \ref{sec:limitations} for more on this issue.}
In the following section, we describe the study design in detail.

\subsection{Method}%

\subsubsection{Research materials}%
We curated a set of prompts and images created with Midjourney, 
a text-to-image generation system and community of AI art practitioners. Using purposeful sampling, we selected 111 images from the corpus of 
thousands of Midjourney images generated by the first author. Our choice to use the author's image corpus has several advantages. The corpus includes images with a range of different prompt modifiers commonly used on Midjourney and we avoid intruding on others' intellectual property rights. Further, the author has experience with text-to-image generation and can distinguish failed attempts from successful ones. This allowed us to purposefully sample images with varying levels of subjective quality. Specifically, we selected 59~images judged as failed attempts and 52 images of high aesthetic quality.
We kept the format of four images per prompt, as it resembles the output a prompt engineer would typically receive on Midjourney.

To assess the aesthetic quality of the 111 images in the dataset, we recruited ten volunteer raters from two academic institutions. The raters had diverse backgrounds in Computer Science, Information Sciences, Human-Computer Interaction, Cognitive Science, Electrical Engineering, and Design. They consisted of 2 Professors, 3 PostDocs, 3 PhD students, 1 Master student, and 1 project engineer (5 men and 5 women, age range 24–48 years). Raters completed a simple binary classification task to classify the images as high or low quality based on their aesthetic appeal. Raters were informed that there was an unequal number of images in each category.
The inter-rater agreement over all images, as measured by Fleiss' kappa, was fair, $\kappa = 0.34$, $z = 23.9$, $p<0.00$, 95\%~CI [0.31, 0.37].%
We discussed the ratings and selected images for further study. Only images with perfect agreement among the ten raters were selected for further study. From the set with perfect agreement among raters, we selected ten high- and ten low-quality images.

We contend that high-quality images can be determined by aesthetic quality ratings from 10 volunteers, without disclosing the prompts to them, for the following reasons.
Classifying images with high and low aesthetics is a relatively easy task~\citep{1606.01621.pdf}.
The evaluation of `high-quality' and `low-quality' images by a limited number of volunteers 
is not aimed at establishing a universal standard of image quality.
We contend that understanding the subjective quality perception of AI-generated images is as important as their fidelity to the prompts.
While we recognize the importance of the alignment between the image and its corresponding prompt in determining quality, our primary focus in this study is on the aesthetic appeal as perceived by individuals without the context of the prompts.
Inspired by the ``wisdom of the crowd'' \citep{surowiecki}, our objective is to leverage perceptual differences in aesthetic appreciation among multiple individuals in order to devise two distinct sets of images, without the influence of the original prompts used to create the images.
Future studies could integrate prompt fidelity as an additional dimension of image quality.

\begin{figure*}[htb]
  \footnotesize
  \centering
  \begin{tabularx}{.9\textwidth}{XXXX}
      \includegraphics[width=\linewidth]{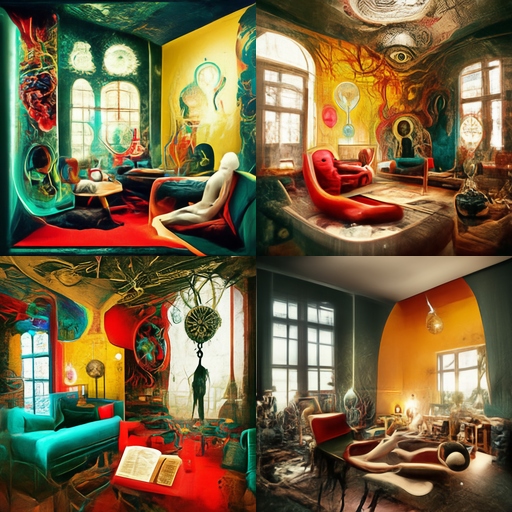}
    & 
      \includegraphics[width=\linewidth]{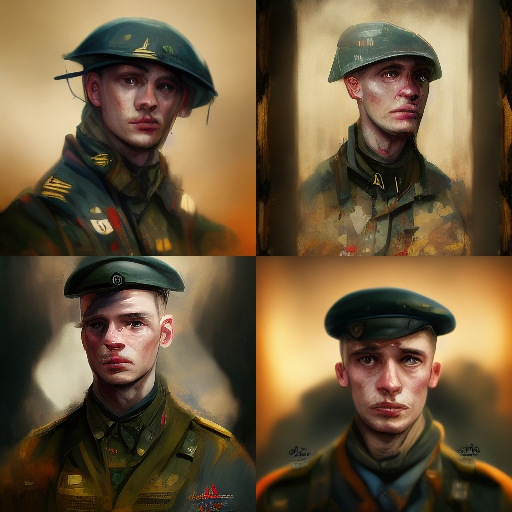}
    &
      \includegraphics[width=\linewidth]{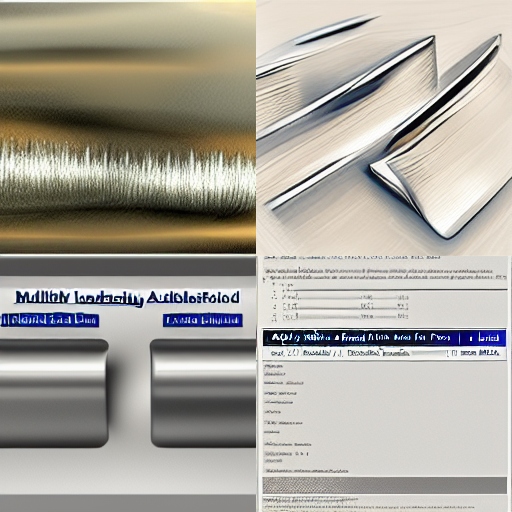}
    &
      \includegraphics[width=\linewidth]{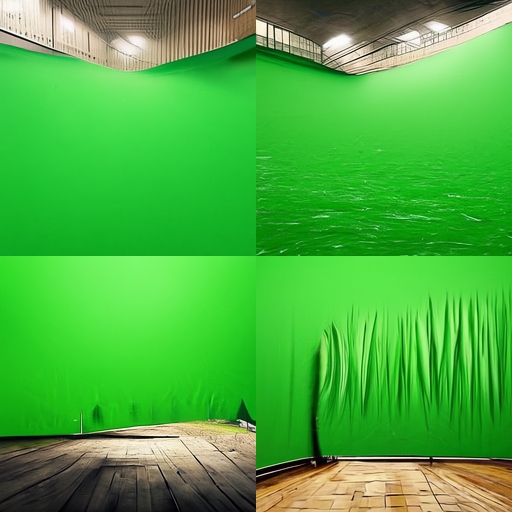}
\\
    \multicolumn{2}{c}{\ref{fig:images}a) High aesthetic appeal}
&
    \multicolumn{2}{c}{\ref{fig:images}b) Low aesthetic appeal}
\end{tabularx}
\caption{Exemplars of images used in Study 2. The full set of images and prompts is listed in Appendix~\ref{appendix:images}.}
\label{fig:images}
\end{figure*}

The final set contains 20~images and respective prompts of varying quality (see Figure~\ref{fig:images} and Appendix~\ref{appendix:images}).

\subsubsection{Study design}%
\begin{figure*}[thb]%
\centering%
\begin{tabularx}{\textwidth}{XX}
    \includegraphics[width=\linewidth]{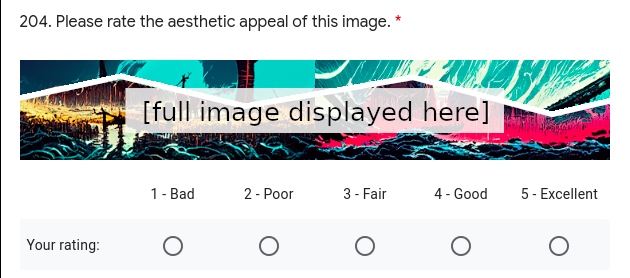}
&
    \includegraphics[width=\linewidth]{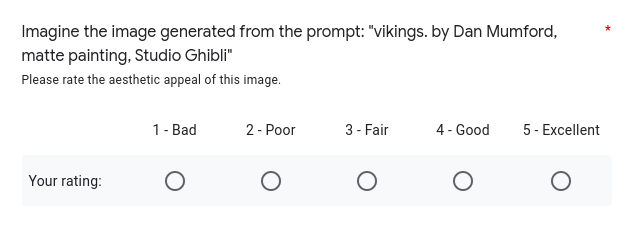}
\\
     \multicolumn{1}{c}{a)} 
&
     \multicolumn{1}{c}{b)} 
\\
\end{tabularx}
  \caption{Example of items used in Study 1 for rating the aesthetic appeal of a) AI-generated artworks and b) prompts. The latter task was prefixed by an instruction to ``imagine the image generated from the prompt.''}%
  \label{fig:promptrating}%
\end{figure*}

We conducted a within-subject experiment with two distinct conditions.
The first condition required participants to rate 20~AI-generated images on a 5-point Absolute Category Rating (ACR) scale \citep{06982326.pdf,06286980.pdf} (refer to Figure~\ref{fig:promptrating}a).
The ACR is an established scale for producing reliable judgments \citep{06982326.pdf},  noted for its insensitivity to variables such as lighting, monitor calibration, language, and country \citep{06286980.pdf}.
In the second condition, participants were presented with 20~textual prompts and asked to imagine and rate the images they believed these prompts would generate, using the same ACR scale.
Here, participants were shown only the prompts, not the actual images.
Each prompt was introduced with ``Imagine the image generated from the prompt: \ldots'' accompanied by a descriptive task reminder (see Figure~\ref{fig:promptrating}b).
%

%

The instructions were carefully designed to avoid confounding factors. For instance, we used neutral wording and avoided referring to the images as artworks to prevent higher positive aesthetic ratings~\citep{10.1163@22134913-00002052.pdf,auteursversie_art_photo_final.pdf,gerger2014.pdf,fpsyg-08-01729.pdf}.
Note that our aim was not to measure exact ground-truth ratings for aesthetic appeal, but to study differences in ratings within participants in a within-subject design.

After the two conditions, we collected basic demographics including participants' self-rated interest in art and experience with practicing art.
We also included an optional open-ended item for participants to elaborate on their experience with text-to-image generation. Experience with art and text-to-image generation were measured on 5-point Likert scales, and self-rated experience with art was measured as a binary variable.

\subsubsection{Participant recruitment and procedure}%
We recruited US-based participants from Amazon Mechanical Turk (MTurk) with a task approval rate greater than 95\% and at least 1000~completed tasks. This combination of qualification criteria is common in crowdsourcing research (e.g., \citep{3491102.3517434}).
The experiment was implemented as a survey task and hosted on Google Forms.
Participants were paid US\$1.50 for completing the survey. The price was determined from the average completion times in a small-scale pilot study ($N=9$, US\$1 per task) \citep{pilotstudies}.%

The task consisted of 31~items in total, including a consent form, an introduction to the study, 20~ratings of prompts, 20 ratings of images, ten demographic items, and one consistency check.
Participants underwent the two conditions (rating of prompt and rating of images) in balanced order. Half of the participants first rated the prompts, then the images, and the other half vice versa.

To prevent bias, we anonymized the filenames of the images and assigned a random numeric code to each image to make it harder to associate the images with the prompts from the previous survey section. 
As a check for consistency, we duplicated one image and collected a rating for this image (L1, see Appendix~\ref{appendix:a2}). Participants who differed in their rating for this duplicated image by greater than one category on the ordinal ACR scale were excluded from analysis.
We excluded four participants for failing this consistency check and another two participants for having completed the survey without completing the task on MTurk.
The final sample included 52 participants.

\subsubsection{Analysis}
We hypothesize that participants can detect a relationship between the quality of a prompt, in terms of its ability to depict visual art through human imagination, and the quality of the visual artwork generated by the text-to-image generation system.
To test this hypothesis, we 
performed a correlation test using Pearson's product-moment correlation to look at the relationship between paired scores for each type.
We investigated the correlation between art experience and average error for each participant.
Average error per participant was calculated by taking the average absolute difference between each pair of prompt and artwork rating.
For example, if all prompts were rated as 2 and all artworks as 5, the average error would be 3.

\subsection{Results}
%
%
\subsubsection{Participants}
Participants ($N=52$) were between 24 and 67 years of age ($M=38.2$~years, $SD=12.98$~years) and included 31~men and 21~women (no non-binary) from diverse educational backgrounds (27~Bachelor's degrees, 10~Master's degrees, among others).
A sizable fraction of the participants (46\%) reported having an educational background in the arts. Twenty-nine participants agreed and nine strongly agreed that they had visited many museums and art galleries  ($M=3.60$, $SD=1.18$). However, participants did not practice art often ($M=3.08$, $SD=1.28$). 
Overall, participants were interested in AI generated art ($M=3.69$, $SD=0.83$), but had little experience with text-to-image generation ($M=2.58$, $SD=1.43$).
Only three participants mentioned having used text-to-image generation (DALL-E mini/Craiyon) before.

\subsubsection{Visual and prompt ratings}
\begin{figure}[htb]%
\centering%
\begin{tabular}{cc}%
  \includegraphics[width=.85\linewidth]{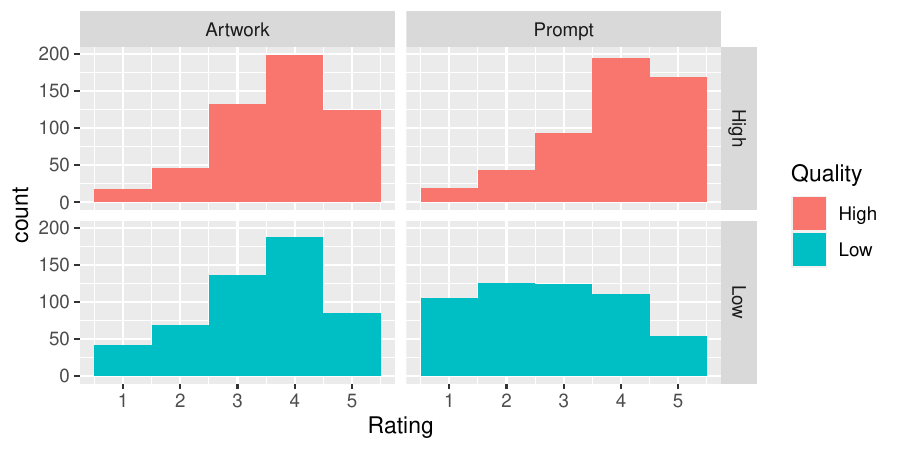}
\end{tabular}
\caption{Histograms of rating scores provided by participants in the two conditions of the within-subject study.
Scores by participants ($N=52$) for images \hl{(left)} and prompts \hl{(right)} with high aesthetic quality (\hl{top;} H1--H10) and low aesthetic quality (\hl{bottom;} L1--L10) in Study 2.}
\label{fig:study2:hist}
\end{figure}

Our study design asked participants to rate both Prompts and Visual artwork. Since participants did not receive special training for prompt engineering or text-based AI art, our goal was to understand the quality of our participants' perceptions. We show the histogram of scores broken into groups for each Art Type (Prompt and Artwork) in Figure~\ref{fig:study2:hist} and    Table~\ref{table:study-2-means} and the Quality (high or low) as described previously. Visually, these show differences across groups, with the distributions of Artworks leaning towards higher quality than prompts.

We used a Kruskal-Wallis rank sum test on these four unique groups, finding a significant difference between the study conditions ($\chi^2=231.4$, $p<10^{15}$, $df=3$).
Following this significant result, we performed post-hoc Dunn's test pairwise across each group with Bonferroni correction for p-values.
Each of these pairs had significant results with a p-value of less than $10^{-4}$, except for Artwork--High versus Prompt--High, in which $p<.004$.
This implies that the median values among all comparisons of groups (i.e. Artwork--High, Artwork--Low, Prompt--High, Prompt--Low) are significantly different from each other.

Participants were able to differentiate images with low visual aesthetic quality from high quality images.
Artwork--High has a higher mean rating ($\mu=3.70$) compared to Prompt--Low ($\mu=3.39$).
Likewise, participants were able to distinguish between high and low quality  by imaging what would be produced based on textual prompts.
Prompt--High has a higher mean rating ($\mu=3.87$) compared to Prompt--Low ($\mu=2.78$).
The overall span between the Artwork High and Low is larger for Prompts ($3.87 - 2.78=1.09$) than for Artworks ($3.70 - 3.39=.31$).
Both High and Low quality Artworks had distributions that favored a rating of 4, while Prompt--Low has a relatively flat distribution across values of 1 to 4 (see Figure~\ref{fig:study2:hist}).%
%
%
%
%
%
%
%
%
%
%
%
%
%
\noindent%
%
%
%
%
\noindent%
\begin{table*}[htb]%
\caption{Average rating across Art Type and Quality in Study 1}.%
\label{table:study-2-means}%
\small
\centering%
\begin{tabular}{llcc}
\toprule
        \textbf{Art Type}	&	\textbf{Quality}	&	\textbf{Mean}	&	\textbf{Std Dev}
        \\	
\midrule
Artwork	&	High	&	3.70	&	1.04	\\
Artwork	&	Low	&	3.39	&	1.15	\\
Prompt	&	High	&	3.87	&	1.07	\\
Prompt	&	Low	&	2.78	&	1.28 \\%
\bottomrule%
\end{tabular}%
\end{table*}%
\subsubsection{Connection between visual image and prompt quality}
While in theory, prompts that can help readers conjure (i.e. visualize or imagine) more aesthetically appealing mental images will also generate better Artwork, it is not clear whether this would be the case for crowdsourced participants.
While our participants were not able to directly associate Prompts to Artworks, each Artwork had a matching Prompt.
We used a Person's product-moment correlation test to measure whether ratings for the Prompt and Artwork are correlated.
The test shows a weak ($r=.29$, 95\% CI [.23, .34]) but significant ($p<10^{-15}$) positive correlation between ratings from Artworks and Prompts.
This indicates that when a Prompt is seen as having a higher quality, that it is also more likely that the Artwork will appear as having a high quality.

\section{Study 2: Writing Prompts}%
\label{sec:study3}%
Our aim with this study is to probe whether laypeople recruited from a crowdsourcing platform have the ability to come up with effective input prompts for text-to-image generation systems with a specific focus on generating digital artworks.
The presence of style modifiers in an input prompt may indicate an understanding of text-to-image generation and how effective prompts can be formulated.

\subsection{Method}%

\subsubsection{Study design}%
We designed a creative crowdsourcing task eliciting three textual prompts from each participant.
The task included a short introduction and the following instructions:
\begin{quote}
    \textit{
    Imagine an artificial intelligence that turns textual input prompts into digital artworks.
    Your task is to produce three artworks.
    To this end, you will write three different input prompts for the artificial intelligence.
    You should aim to maximize the visual attractiveness and aesthetic qualities of the digital artworks generated from your input prompts.}
\end{quote}

Participants were asked to make their artworks as visually attractive and high-quality as possible.
We did not mention that prompt modifiers could be used in the prompt and wrote the instructions to avoid priming participants with a specific style (i.e., we told participants to produce `artworks' rather than `paintings').
Note, however, that we did not aim to precisely measure attractiveness and quality, but wanted participants to think about the overall visual and aesthetic quality of the images. Participants were told that there was no right or wrong answer, but tasks would be rejected if they didn't follow the instructions.
To avoid influencing participants prompt modifications in our follow-up study (Study 3), we merely collected prompts from participants in this study --- the outcome of the image generation was not shown to participants in this study.

As additional questions in the task, we asked whether the participant had experience with text-to-image generation and we collected basic demographics.
Participants were paid US\$0.16 per completed task. The pricing was estimated from the average task completion times in a pilot study ($N=10$, US\$0.12 per task).
In this pilot study, we noticed some participants wrote a series of consecutive instructions for the AI. The task design and instructions were subsequently adjusted to elicit complete prompts.

\subsubsection{Participant recruitment}%
We recruited 137~unique participants recruited from Amazon Mechanical Turk using the same qualification criteria as in Study~1.
Ten tasks had to be rejected due to clearly no attempt being made to answer the task with relevant information. The ten tasks were republished for other participants. After collecting the data, we manually reviewed the results and removed a further twelve responses from participants who obviously tried to game the task.
The final set includes 375~prompts written by 125~unique participants (three prompts per participant).

\subsubsection{Analysis}%
The analysis of the prompts was conducted with mixed methods.
For each prompt, we qualitatively and quantitatively analyzed the prompts, as follows.

\paragraph{Prompt modifiers}%
Our analysis focused on identifying the presence of specific keywords and phrases frequently used within the AI art community to influence the style and quality of AI-generated images \citep{modifiers,aiartcreativity}. We opted for manual analysis in this case. An initial screening indicated that a very small portion of the prompts utilized prompt modifiers, making automated methods unnecessary for this specific task.

The analysis process involved the first author of this paper who systematically reviewed each prompt, with focus on identifying specific language patterns and stylistic phrases known to impact the AI's generative capabilities. This included looking for explicit instructions or adjectives that might alter the style or quality of the generated images.
Given the clear and specific nature of these prompt modifiers, the coding focused on the presence or absence of these elements in each prompt.
We determined that the coding process was straightforward -- i.e., it
did not require complex categorization or subjective interpretation -- and, thus, did not necessitate validation via inter-rater agreement \citep{McDonald_Reliability_CSCW19.pdf}.

We analyzed whether the prompts contained certain keywords and phrases commonly used in the AI art community to modify the style and quality of AI generated images \citep{modifiers,aiartcreativity}.
We decided on manual analysis because a preliminary screening revealed that very few prompts contained prompt modifiers.
Each prompt was analyzed by an author of this paper. We coded the presence of prompt modifiers and report on their nature and use.
We did not calculate inter-rater agreement because the coding was straight-forward~\citep{McDonald_Reliability_CSCW19.pdf}.%

\paragraph{Descriptive language}%
A prompt written in descriptive language is likely to generate images of high quality.
We quantitatively assessed whether the prompts contained descriptive language by calculating a number of statistical indices for each prompt:
\begin{itemize}
    \item The 
    number of words (tokens) and unique words (types) in the prompt.
        In general, longer prompts are more likely to include certain keywords
        (whether on purpose or by accident)
        that may trigger the image generation system to generate images with high quality or in a certain style.
    \item The Type-Token Ratio (TTR) \citep{TTR}, a standard measure for lexical diversity defined as the number of types divided by the number of tokens in the prompt.\footnote{We also experimented with other indices of lexical diversity, such as the Moving-Average Type-Token Ratio (MATTR) \citep{covington2010.pdf} and
    the Measure of Textual Lexical Diversity (MTLD) \citep{MTLD_MA_Wrap}.
    However, these measures highly depend on the text length \citep{30200474.pdf}.
    Only a small fraction of the prompts in our sample meet the recommended minimum number of tokens for applying lexical diversity measures~\citep{1-s2.0-S1075293520300660-main.pdf}.
    The use of lexical diversity indices, such as the TTR, for comparing texts of different size is not recommended~\citep{30200474.pdf}.
    In our study, we do not use the TTR for comparing the lexical diversity of prompts, but to assess the amount of repetition of tokens in the prompt.}
    A token, in this case, is a discrete word whereas a type is a unique token in the prompt.
    For calculating the TTR, we used Kristopher Kyle's lexical-diversity Python package \citep{pythonpackage}.%
\end{itemize}%
\subsection{Results}%
\label{sec:study2:results}
\subsubsection{Participants}%
The 125 participants in our sample included 55 men, 67 women, 1 non-binary, and two participants who did not to disclose their gender identity.
The age of participants ranged from 19 to 71 years ($M=41.08$ years, $SD = 13.44$ years).
The majority of participants (98.40\%) reported English being their first language.
Thirty-seven participants (30.33\%) responded positively to the question that they had \textit{``experience with text-based image generation systems.''} We had no explanation for this surprisingly high number at this point, but inquired more about the participants' background in our follow-up study in Section~\ref{sec:study4}.
Median completion times were higher than estimated in the pilot study, reaching 197 seconds. It is possible that completion times are skewed due to participants reserving tasks in bulk.

\subsubsection{On the use of descriptive language}
\label{sec:study3:descriptivelanguage}
The prompts were of varying length, ranging from 1 to 134 tokens with an average of 12.54~tokens per prompt ($SD=14.65$ tokens).
Overall, the length of prompts was appropriate for text-to-image generation with only four participants producing overly long prompts.
%
%
On average, participants used 3.27 nouns to describe the subjects in their prompt ($SD=3.36$).
Participants used verbs only sparingly in their prompts ($M=0.36$, $SD=1.02$).
The average number of prepositions ($M=1.78$, $SD=2.27$) was higher than the average number of adjectives ($M=1.65$, $SD=1.95$). However, this number is skewed by four participants who provided long prompts.
These participants were very specific in what their images should contain, with many prepositions being used to denote the relative positions of subjects in the artwork ($Max=21$ prepositions per prompt).

Overall, participants used rich descriptive language.
The participants were creative and often described beautiful natural scenery.
The main topics in the participants' prompts were landscapes, sunsets, and animals. 
We note that the richness of the language in the prompts primarily is a result of the use of adjectives.
On average, participants used 1.65~adjectives in their prompt ($SD=1.95$).
Colors, in particular, were popular among participants to describe the subjects in their artworks.
The following prompts exemplify the creativity and the use of descriptive language among participants:
\begin{itemize}%
    \item
    \textit{%
    beautiful landscape with majestic mountains and a bright blue lake
    }
    \item
    \textit{%
    bright yellow sun against a blue sky with puffy clouds
    }
    \item
    \textit{%
    A fruit bowl with vibrant colored fruits in it and a contrasting
    background
    }
    \item
    \textit{%
    Dragon on the tower of a castle in a storm.
    }
    \item
    \textit{%
    Knight holding a sword that shines in the sunlight
    }
    \item
    \textit{%
    A white fluffy puppy is playing in the grass with a large blue ball that is twice his size.
    }
    \item
    \textit{%
    A shiny black horse with eyes like coal run in a lush green grassy field
    }
    \item
    \textit{%
    There should be a beautiful green forest, full of leaves, with dark brown earth beneath, and a girl in a dress sitting on the ground holding a book.%
    }%
\end{itemize}%

More than half of the prompts (58.13\%) did not repeat any tokens (that is, they had a TTR of 1; $M=0.94$, $SD=0.10$). Most of the repetitions in prompts stem from the participants' need to identify the relative positions of subjects in the image (e.g., \textit{``[...] Touching the black line and going all the way across the top of the black line should be a dark green line. Above the dark green line should be a medium green line. [...]''}).
Repetitions, as a stylistic element in prompts \citep{modifiers}, were not being used.
Only 27~prompts (7.2\% of all prompts) contained cardinal numbers ($M=0.07$, $SD=0.29$).
    Two cardinal numbers referred to a period in time which could potentially trigger the image generation system to produce images in a certain style.

Even though we tried to mitigate it in the task design and the instructions, we noticed 18~participants (14.4\%) still provided direct instructions to the AI instead of prompts describing the image content.
    These participants either wrote three separate instructions to the AI (e.g., \textit{``Generate a white 250 ml tea glass [...],''} \textit{``Draw three separate triangles [...],''} and \textit{``Show me some digital artwork from a brand new artist.''}) or they wrote three consecutive instructions as we had observed in our pilot study. The latter may not include nouns as subject terms and could thus result in images with an undetermined subject (e.g., \textit{``sharpen image''}).
Two participants thought they could chat with the AI, asking it, for instance, \textit{``Which do you prefer: starry night sky or blue sea at dawn?,''} \textit{``Enter your favorite geometric shape,''} and \textit{``Can you paint me a rendition of the Monalisa?''}.%


\subsubsection{On the use of prompt modifiers}%
\label{sec:study3:modifiers}%
Even though participants were specifically instructed to create a digital artwork, we found only very few participants included style information in their prompts.
Many participants described a scene in rich descriptive language, but neither mentioned artistic styles, artist names, genres, art media, nor specific artistic techniques.
The participants' prompts may have described an artwork, but without style information, the style of the generated image is left to chance and the resulting images may not match the participants' intent and expectations.

Overall, the prompts did not follow the prompt template mentioned in Section~\ref{sec:promptmodifiers} and best practices common in the AI art community were not followed.
Only one participant made purposeful use of a prompt modifier commonly used in the AI art community. This prompt modifier is \textit{``unreal engine.''}\footnote{The long-form of this modifier is \textit{``rendered in UnrealEngine,''} a computer graphics game engine. Images generated with this prompt modifier may exhibit increased quality due to photo-realistic rendering.}
The participant used this modifier in all her three prompts by concatenating it to the prompt with a plus sign, e.g.%
    \textit{``rainbow tyrannosaurus rex + unreal engine.''}
A small minority of participants used generic keywords that could trigger a specific style in text-to-image generation systems.
    For instance, the generic term \textit{``artwork''} was used in 16~prompts (4.3\%).
The following list of examples reflects almost the entire set of prompts containing explicit style information among the 375 prompts written by participants (with style modifiers underlined):%
\begin{itemize}%
\item
        \underline{Cubism portrait} of a Labrador Retriever using reds and oranges
\item
        \underline{Paint} a \underline{portrait} of an old man in a park.
\item
        \underline{Draw} a \underline{sketch} of an airplane.
        \vspace{.2\baselineskip}
\item
        \underline{Abstract} trippy colorful background
        \vspace{.2\baselineskip}
\item
        \underline{surreal} sky castle
\item
        Can you \underline{paint} me a rendition of the \underline{Monalisa}?
        \vspace{.2\baselineskip}
\item
        \underline{Bob Ross}, \underline{Claude Monet}, \underline{Vincent Van Gogh}
        \vspace{.2\baselineskip}
\item
        Are you able to produce any of \underline{rodans work}.
        \vspace{.2\baselineskip}
\item
        what can you do, can you make \underline{pointillism artwork}?
\end{itemize}%
Besides this sparse --- and sometimes accidental --- addition of style information, we find that overall, participants did not control the style of their creations. 
Instead of prompt modifiers, the participants' artwork styles were mainly determined by the participants' use of descriptive language.
%
%
%
%

\section{Study 3: Improving Prompts}%
\label{sec:study4}%
In a follow-up study, we investigated whether participants could improve their artworks.
This study aimed to answer the question of whether prompt engineering is an innate skill that we humans apply intuitively or whether it is an acquired skill that requires expertise and practice (e.g., via learning to write prompts from repeated interactions with the text-to-image generation system) and knowledge of certain keywords and key phrases (prompt modifiers), as discussed in Section \ref{sec:promptmodifiers} and Section \ref{sec:study3:modifiers}.
%
We hypothesize that if prompt engineering is a learned skill, participants will not be able to significantly improve their artworks after only one iteration.
%
%
\subsection{Method}%

\subsubsection{Study design}%
We invited the same participants who participated in Study~2 to review images generated from their own prompts.
    Participants were then asked to improve their three prompts.
To this end, we designed a task that introduced the participant to the study's purpose, using the same instructions as in the previous study.
We additionally highlighted that if the images presented to the participant did not look like artworks, the prompt should be adjusted. 
Like in the previous study, we avoided to mention that prompt modifiers could be used to achieve this aim.

Participants were given five images for each of the three prompts they wrote in Study 2. We used the workerId variable on MTurk to load the participant's previous prompts and images. Participants were then asked to rewrite and improve their three prompts. The task included two input fields, one pre-filled with their previous prompt and one for optional negative terms. In practice, negative terms are an important part of the toolbox of prompt engineers, primarily used for controlling the subject and quality of the image generation \citep{modifiers}. For example, defining \textit{``watermark''} and \textit{``shutterstock''} as negative terms can reduce the occurrence of text and watermarks in the resulting image.
Given the task of improving their previously generated artworks, we introduced negative terms as a possible tool for making improvements in Study 3.
We studied this by incorporating it into our study design. Participants were introduced to the potentially surprising effects of negative terms with the help of an example. The example explained that adding \textit{``zebra''} as a negative term to a prompt for a pedestrian crossing could potentially result in an image of a plain road (due to stripes being removed).

For each prompt, we also collected information on whether the images matched the participant's original expectations (given the previous prompt)
and whether the participant thought the prompt needed improvement (both on a Likert-scale from 1~--  Strongly Disagree to 5~-- Strongly Agree).
    The latter was added to identify cases in which participants thought that no further improvement of the prompt was necessary.
We also asked participants to rate their confidence that the new prompt would result in a better artwork  (on a Likert scale from 1~-- Not At All Confident to 5~-- Highly Confident).
The task concluded with demographic questions, including
the participant's experience with text-based image generation and interest in viewing and practicing art.
The task design was tested and improved in a small-scale pilot study ($N=8$; US\$1 per task).
    The payment was set to US\$1.75, aiming for an hourly pay of above minimum wage in the United States.

\subsubsection{Research materials}%
\label{sec:study4-materials}%
In this section, we describe how we selected an image generation system and how we generated images from the participants' prompts.%
\paragraph{System selection}%
We experimented with different text-to-image generation systems, including
    CLIP Guided Diffusion (512x512, Secondary Model)\footnote{https://colab.research.google.com/drive/1mpkrhOjoyzPeSWy2r7T8EYRaU7amYOOi},
    CLIP Guided Diffusion (HQ 512x512 Uncond)\footnote{https://colab.research.google.com/drive/1QBsaDAZv8np29FPbvjffbE1eytoJcsgA},
    DALLE-E mini\footnote{https://github.com/borisdayma/dalle-mini},
    Disco Diffusion 5.3 and 5.4\footnote{https://github.com/alembics/disco-diffusion},
    Latent Diffusion\footnote{https://github.com/CompVis/latent-diffusion},
    and Majesty Diffusion 1.3\footnote{https://github.com/multimodalart/majesty-diffusion}.
In the end, we selected Latent Diffusion for two main reasons. Latent Diffusion is the foundation for many of the community-driven adaptations and modifications.
More importantly, the system is deterministic and leads to reproducible outcomes. Consecutive runs with the same seed value will generate the same images. This is a crucial requirement since we aim to compare images in between studies.
\paragraph{Image generation}%
We generated images for the participants' prompts with Latent Diffusion using the following configuration settings:
    text2img-large model (1.4B parameters),
    seed value 1040790415,
    eta 1.0,
    ddim steps 100, and
    scale 5.0.
%
Even though the system is capable of generating images at higher resolutions, we decided to generate images of $256\times256$ pixels to avoid the quirks that often occur when generating images in resolutions that the model was not trained on. 
The image generation job yielded 1875 images (125 participants $\times$ 3 prompts per participant $\times$ 5 images per prompt).
After collecting the revised prompts from participants, we generated another set of 1875 images using the same seed value and configuration settings as before. Negative terms were used in this second set, if provided by the participant.%

Some hand-selected images generated from the prompts are depicted in Figure~\ref{fig:study34-examples}.
Many images were of photo-realistic quality, depicting landscapes, sunsets, beaches, and animals. Besides photographs, artistic styles included paintings, graphic designs, abstract artworks, as well as magazine and book covers.
Some images contained text and many images contained watermarks.%
%
\noindent%
{%
\setlength\tabcolsep{2pt}%
\begin{figure}[thb]%
\centering%
\begin{subfigure}[b]{\textwidth}
\centering%
\begin{tabularx}{.95\textwidth}{XXXXXXX}%
  \includegraphics[width=\linewidth]{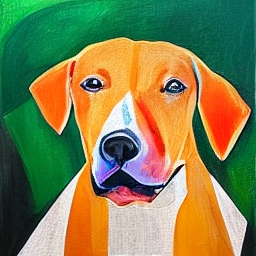}
&
  \includegraphics[width=\linewidth]{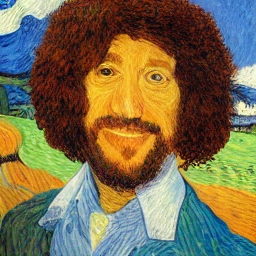}
&
  \includegraphics[width=\linewidth]{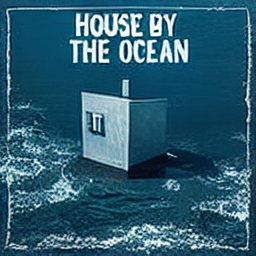}
&
  \includegraphics[width=\linewidth]{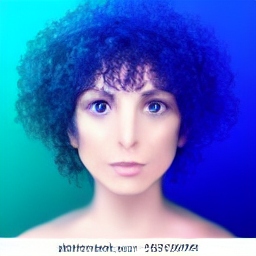}
&
  \includegraphics[width=\linewidth]{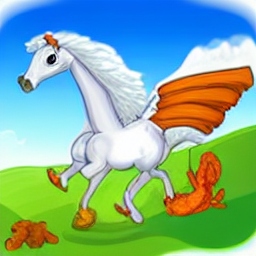}
&
  \includegraphics[width=\linewidth]{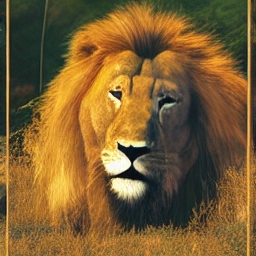}
&
  \includegraphics[width=\linewidth]{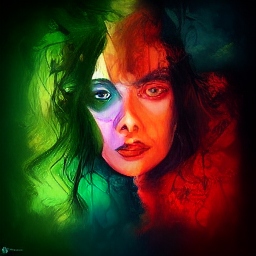}
\\
  \includegraphics[width=\linewidth]{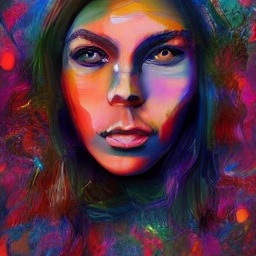}
&
  \includegraphics[width=\linewidth]{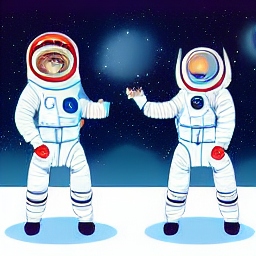}
&
  \includegraphics[width=\linewidth]{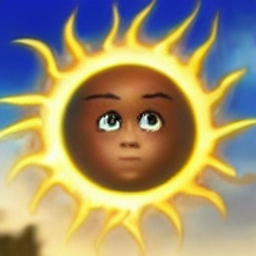}
&
  \includegraphics[width=\linewidth]{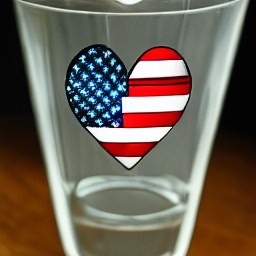}
&
  \includegraphics[width=\linewidth]{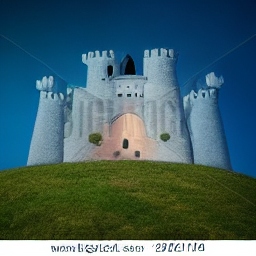}
&
  \includegraphics[width=\linewidth]{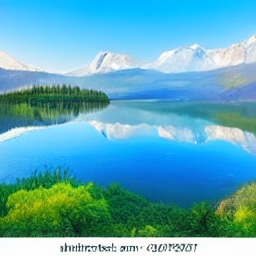}
  &
  \includegraphics[width=\linewidth]{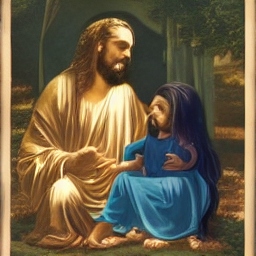}
\\
  \includegraphics[width=\linewidth]{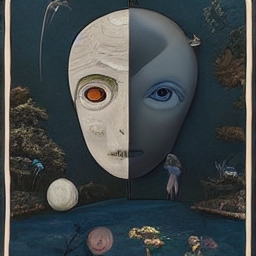}
&
  \includegraphics[width=\linewidth]{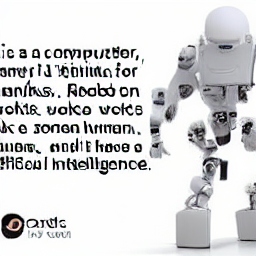}
&
  \includegraphics[width=\linewidth]{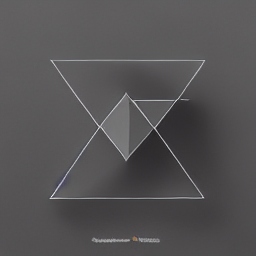}
&
  \includegraphics[width=\linewidth]{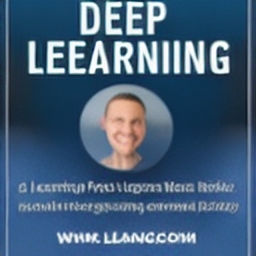}
&
  \includegraphics[width=\linewidth]{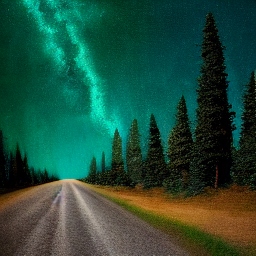}
&
  \includegraphics[width=\linewidth]{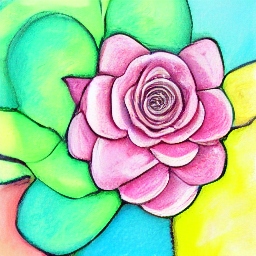}
&
  \includegraphics[width=\linewidth]{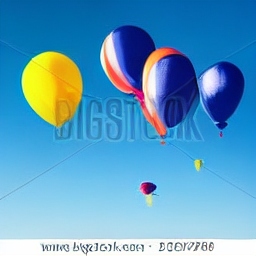}
\\
\end{tabularx}%
    \subcaption{}
    \label{fig:study34-examples:a}%
\end{subfigure}%
\vspace{.5\baselineskip}
\\
\begin{subfigure}[b]{\textwidth}%
\centering%
\begin{tabularx}{.95\textwidth}{XXXXXXX}%
  \includegraphics[width=\linewidth]{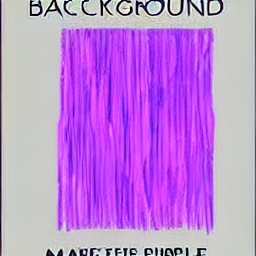}
&
  \includegraphics[width=\linewidth]{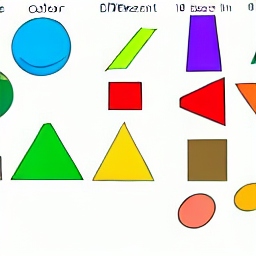}
&
  \includegraphics[width=\linewidth]{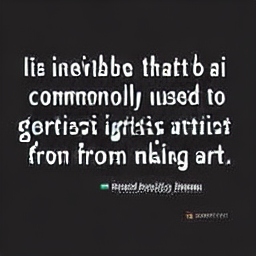}
&
  \includegraphics[width=\linewidth]{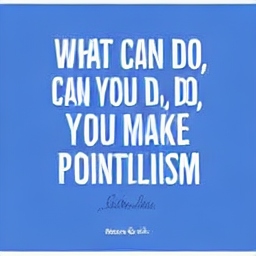}
&
  \includegraphics[width=\linewidth]{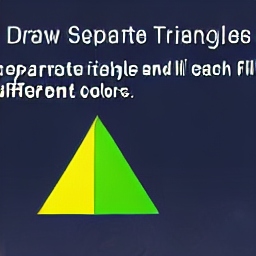}
&
  \includegraphics[width=\linewidth]{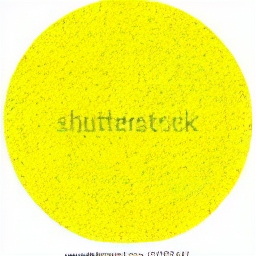}
&
  \includegraphics[width=\linewidth]{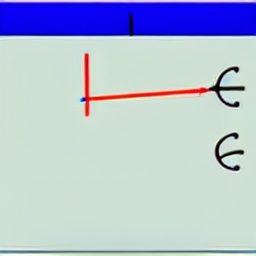}
\\
  \includegraphics[width=\linewidth]{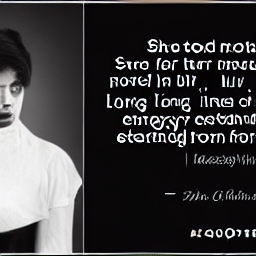}
&
  \includegraphics[width=\linewidth]{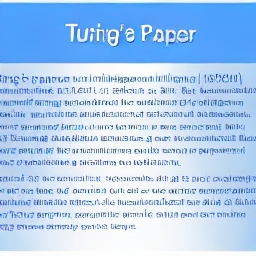}
&
  \includegraphics[width=\linewidth]{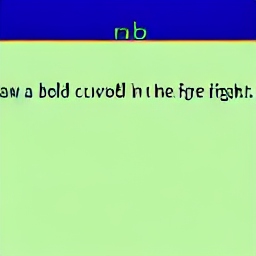}
&
  \includegraphics[width=\linewidth]{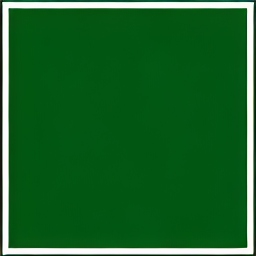}
&
  \includegraphics[width=\linewidth]{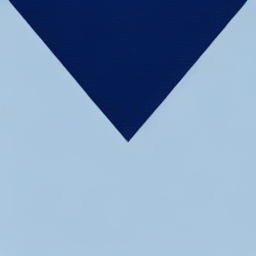}
&
  \includegraphics[width=\linewidth]{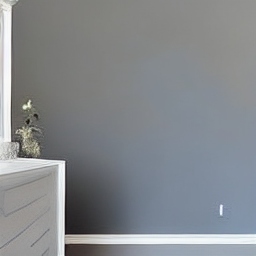}
  &
  \includegraphics[width=\linewidth]{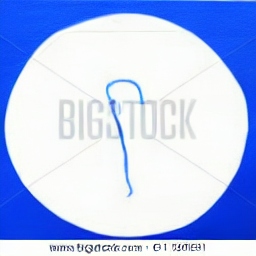}
\\
\end{tabularx}%
    \subcaption{}
    \label{fig:study34-examples:b}%
\end{subfigure}%
\caption{Selected exemplars of a) successful and b) failed image generations from worker-provided prompts. The images in Figure~\ref{fig:study34-examples:a} were selected to represent a variety of different styles and are not representative of the whole set of images. The images in Figure~\ref{fig:study34-examples:b} depict some of the recurring issues in images generated from worker-provided prompts.}%
\label{fig:study34-examples}%
\end{figure}%
}%
%

\subsubsection{Analysis}%
\label{sec:study4:analysis}%
We analyzed the two sets of prompts and images written in studies~2 and~3 as follows.

\paragraph{Analysis of prompts}
To measure the amount of changes in the prompts,
we calculated the number of tokens added and removed using parts-of-speech tagging as well as the Levenshtein distance~\citep{Levenshtein1966a.pdf}, a measure of lexical similarity denoting the minimum number of edits needed to change one string into another.
The Levenshtein distance provides a simple measure for us to describe how much a prompt has changed in between the two studies. This is important to know, because it also reflects the participant's satisfaction with the generated output. Clearly, there will be satisficers and maximizers among participants \citep{maximizing.pdf}. Some participants spend more time on prompt writing than others. We argue it is important to have a notion of the change in between prompts, measured by the Levenshtein distance.
To understand the nature of the changes, the first author inductively developed a coding scheme \citep{hsieh2005three} with eight categories: adjectives/adverbs, subjects, prepositions, paraphrasing/synonyms, reordering, cardinal numbers, simplification, and presence of prompt modifiers.
After discussing the codes among all authors and revising the codes, the first author coded all prompts and generated a co-occurrence matrix of changes made by participants. Note that we understand ``subjects'' in the sense of subject terms~\citep{modifiers} for image generation (e.g. ``a woman holding a phone'' would have two subjects (woman and phone). Synonyms were analyzed at the level of individual words and parts of sentences.

\paragraph{Analysis of the revised images}%
We evaluated the images according to the following process, developed collaboratively by the authors.
Initially, a spreadsheet was created with the two sets of prompts and their respective five images from Studies~2 and~3. Through a detailed discussion of 30~image-text pairs, the authors developed a set of evaluation criteria grounded in both the study's objectives and relevant literature. Similar criteria have been used in image evaluation studies, such as \citep{2309.15807.pdf}.
This approach ensured a balance between the specificity of the study and established methodologies in image evaluation. The criteria were designed to encompass both objective and subjective aspects of the images, including binary categories for failed image generations, the extent of style and subject change, and improvements in consistency. Additionally, we included ratings for details, contrast, color, distortions, watermarks, and an overall subjective impression of quality.
We acknowledge that while some elements of the evaluation were inherently subjective, they were rooted
in discussions among the authors to ensure they were relevant and appropriate for the context of this study. We aimed to create a comprehensive evaluation framework that not only aligns with existing standards but also caters to the unique aspects of AI-generated imagery.

Using the evaluation criteria, each author then individually rated 50 pairs of images along these criteria.
After this initial round of coding, the authors discussed the results and decided to add four more criteria to the coding scheme. The final set of criteria included binary categories for failed generations, amount of style and subject change, and whether consistency improved, as well as ratings for details, contrast, color, distortions, watermarks, and overall subjective impression of quality. After a second round of coding, the authors cross-checked their evaluations and resolved differences through discussion.

\subsection{Results}
\subsubsection{Participants}
The sample consisted of 50~participants (40\% of the participants who participated in Study~2).
Participants included 25~men, 24~women, and 1~person who preferred not to disclose the gender identity, aged 20 to 71 years ($M=42.76$, $SD=14.63$).
Participants came from varied educational backgrounds, including some completed college courses (17 participants), Bachelor's degrees (22 participants), Master's degrees (4 participants), and doctorate degrees (2 participants).
%
%
\noindent%
\begin{figure}[htb]%
\centering%
\includegraphics[width=.75\linewidth]{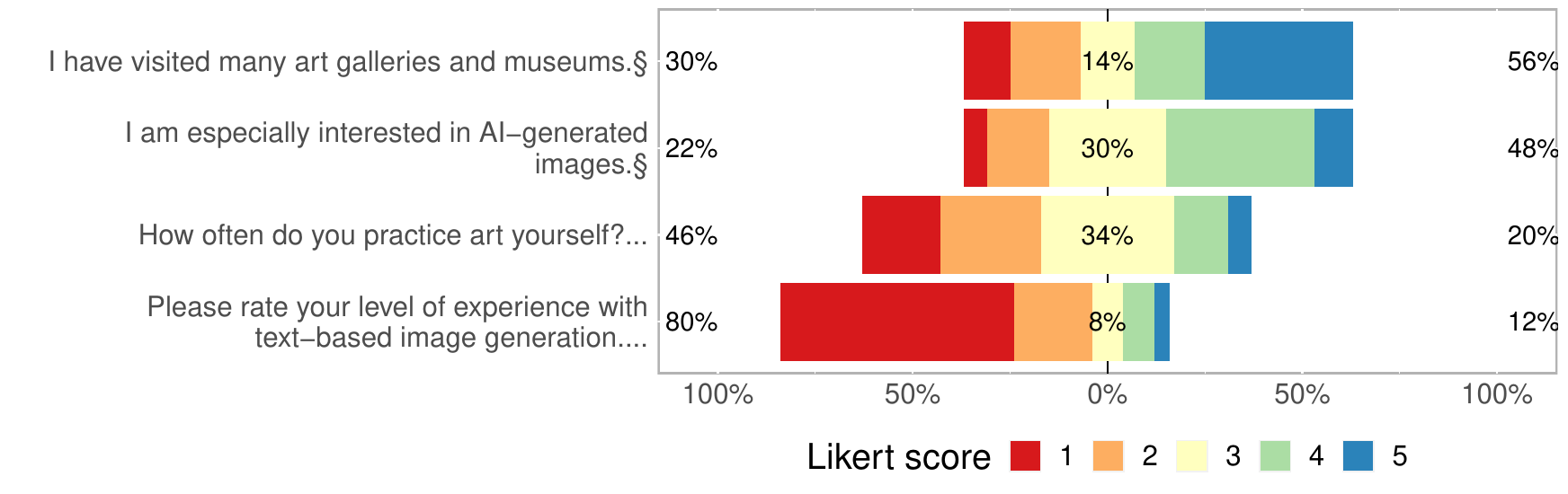}%
\\
\footnotesize{%
\raggedright
§ From 1 -- Strongly Disagree to 5 -- Strongly Agree\\
‡ 1 -- Never, 2 -- Rarely, 3 -- Sometimes, 4 -- Often, 5 -- Very often\\
† 1 -- Not at all experienced, 2 -- Slightly experienced, 3 -- Moderately experienced, 4 -- Very experienced, 5 -- Extremely experienced\\
}%
\caption{Background of the crowd workers participating in Study 3.}%
\label{fig:study4:likerts}%
\end{figure}%
%
Seven out of ten participants had an educational background in the arts. Some participants were interested in visiting museums and AI-generated imagery, but most did not practice art themselves and 80\% had little or no experience with text-to-image generation.

Approximately 40\% of participants were disappointed with the generated images, while 55\% of participants' expectations were met. Around 60\% of participants believed the images needed improvement, and a similar percentage of participants were confident that their revised prompts would improve the generated images.

\subsubsection{Participants' revised prompts}%
\begin{figure}[htb]%
\centering%
\begin{subfigure}[b]{0.42\textwidth}
    \includegraphics[width=.905\linewidth]{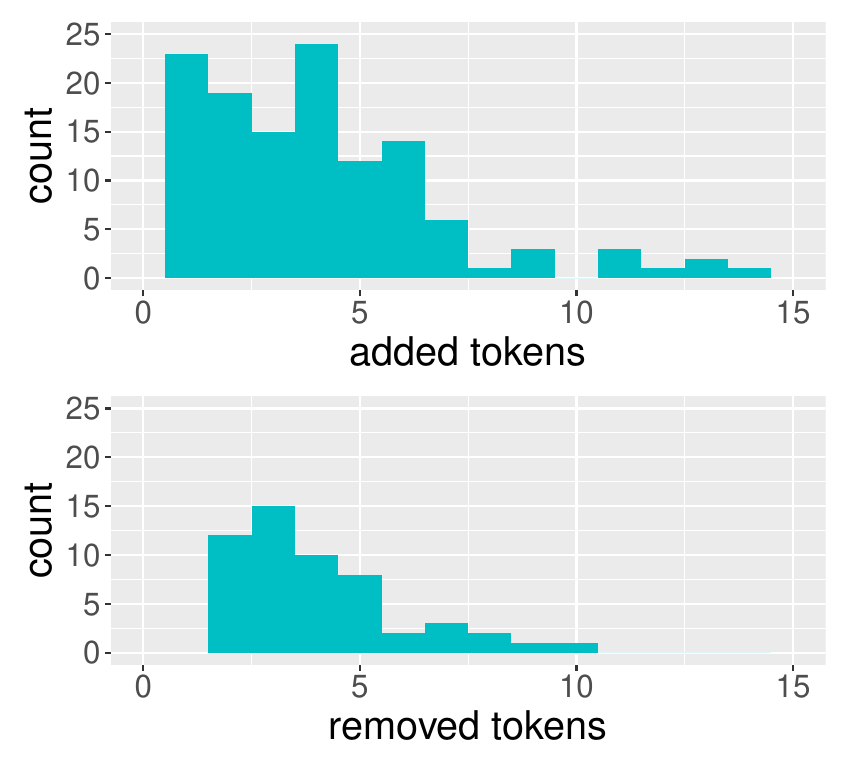}%
    \subcaption{%
        \centering
        Histogram of changes in tokens
    }%
    \label{fig:study4:changes:histograms}%
\end{subfigure}%
\begin{subfigure}[b]{0.42\textwidth}%
    \includegraphics[width=\linewidth]{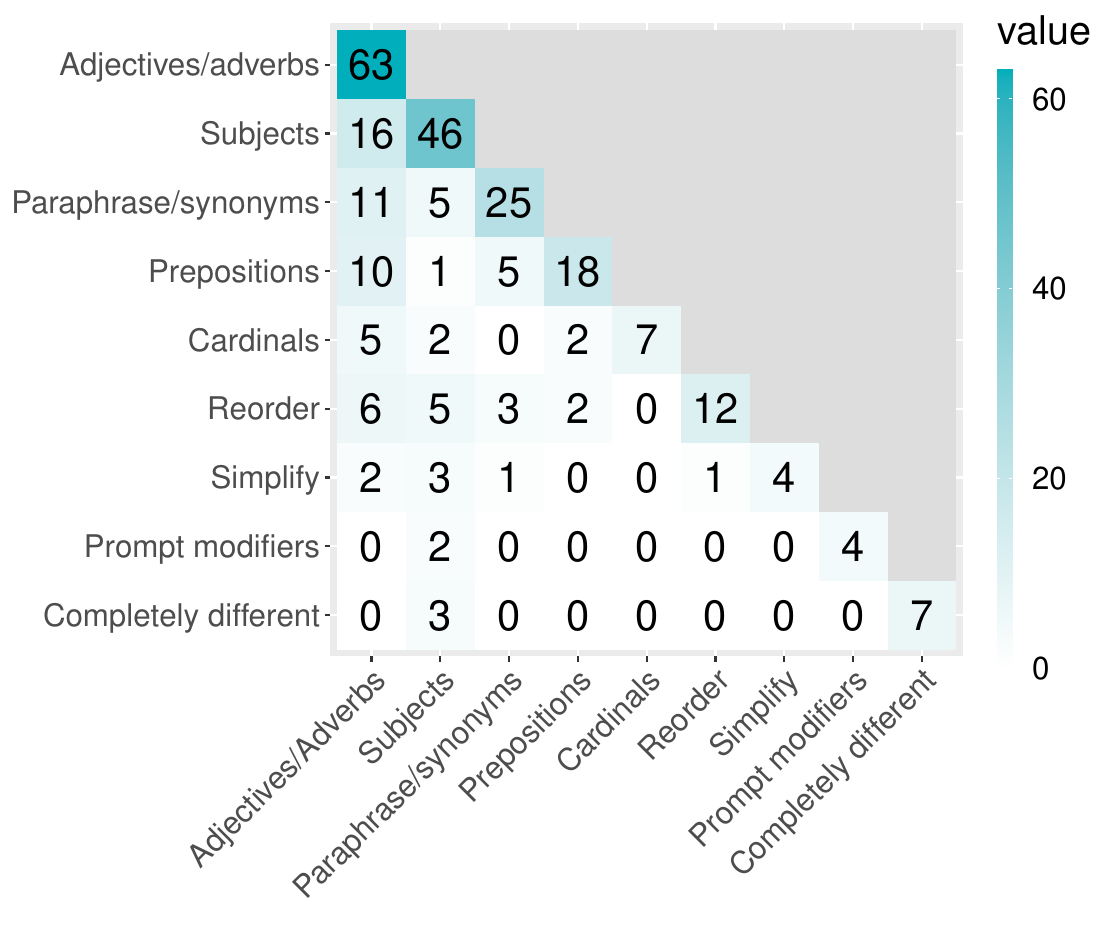}%
    \subcaption{%
        \centering
        Co-occurrence matrix of changes
    }%
    \label{fig:study4:changes:cooccurence}%
\end{subfigure}%
\caption{\hl{Participants} added more tokens than they removed in Study 3 (Figure~\ref{fig:study4:changes:histograms}).
Figure~\ref{fig:study4:changes:cooccurence} depicts the changes that often co-occurred with one another. For instance, a change (addition or removal) to an adjective often co-occurred with changes to other adjectives in the prompt.}
\label{fig:study34:changes}%
\end{figure}

The average Levenshtein distance between the participants' two prompts (not including negative terms) was 28.1 ($SD=25.0$).
A computational analysis of the changes with parts-of-speech tagging shows that participants added over twice as many tokens as they removed~--- 538 added tokens versus 243 removed tokens (see Figure~\ref{fig:study4:changes:histograms}).
Nouns were added most often (29.55\% of added tokens), followed by adjectives (22.12\%), prepositions (17.84\%) and determiners (8.55\%).
The same types of tokens were also most often removed (28.81\% of removed tokens were nouns, 16.87\% prepositions, 13.17\% adjectives, and 8.64\% determiners).
In 11~prompts (7.33\%), the participant neither changed the prompt nor provided a negative term.
    Six of these instances consisted of participants pasting random snippets of text.

%
%
\begin{table*}[thb]%
\small%
\caption{Evaluation of changes in the two sets of images generated \hl{from prompts} in Study~3.}%
\label{tab:imageevaluation}%
\centering
\begin{tabular}{rccccccc}%
    \toprule%
     & details &	contrast & color &	distortions &	watermarks &			consistency &  overall
    \\%
\midrule%
    worse  & 17 (11.3\%) & 17 (11.3\%) & 12 (8.0\%) & 32 (21.3\%) & 31 (20.7\%) & 23 (15.3\%) & 23 (15.3\%) \\
    same   & 81 (54.0\%) & 85 (56.7\%) & 88 (58.7\%) & 99 (66.0\%) & 85 (56.7\%) & 95 (63.3\%) & 77 (51.3\%) \\
    better & 52 (34.7\%) & 48 (32.0\%) & 50 (33.3\%) & 19 (12.7\%) & 34 (22.7\%) & 32 (21.3\%) & 50 (33.3\%) \\
\bottomrule%
\end{tabular}%
\end{table*}%
\vspace{0pt}%
%

%
\newlength{\thumbnailwidth}%
\setlength{\thumbnailwidth}{1.5cm}%
\newlength{\skiplength}%
\setlength{\skiplength}{4pt}%
\noindent%
{%
\begin{figure}[thb]%
\centering%
\begin{subfigure}[b]{0.32\textwidth}
\noindent
    \begin{tabularx}{\textwidth}{cX}%
      \includegraphics[width=\thumbnailwidth,valign=t]{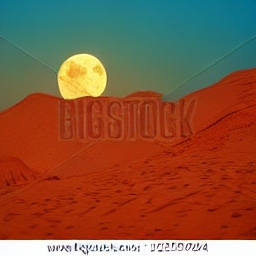}%
        \vspace{\skiplength}
        &
        \scriptsize
        bright full moon just rising over the desert	
        \\
      \includegraphics[width=\thumbnailwidth,valign=t]{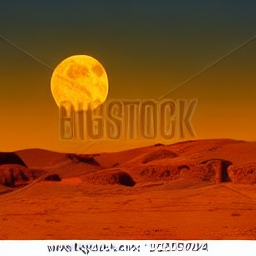}%
        &
        \scriptsize
        bright full \textbf{amber colored} moon just rising over the desert	
    \end{tabularx}%
    \subcaption{}%
    \label{fig:study4:examples:a}
\end{subfigure}
\begin{subfigure}[b]{0.32\textwidth}
    \begin{tabularx}{\textwidth}{cX}%
      \includegraphics[width=\thumbnailwidth,valign=t]{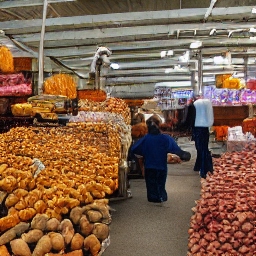}%
        \vspace{\skiplength}
        &
        \scriptsize
        A famers market in Nebraska in the early fall.	
        \\
      \includegraphics[width=\thumbnailwidth,valign=t]{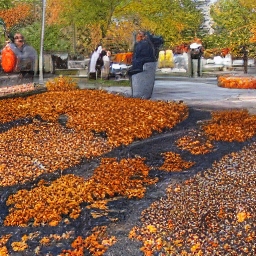}%
        &
        \scriptsize
        An \textbf{outdoor} famers market in Nebraska in the early fall.	
    \end{tabularx}%
    \subcaption{}%
    \label{fig:study4:examples:b}
\end{subfigure}
\begin{subfigure}[b]{0.32\textwidth}
    \begin{tabularx}{\textwidth}{cX}%
      \includegraphics[width=\thumbnailwidth,valign=t]{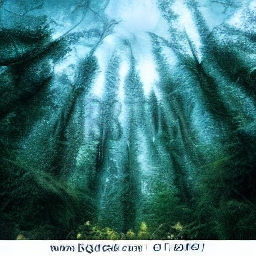}%
        \vspace{\skiplength}
        &
        \scriptsize
        A forest with scary trees all around
        \\
      \includegraphics[width=\thumbnailwidth,valign=t]{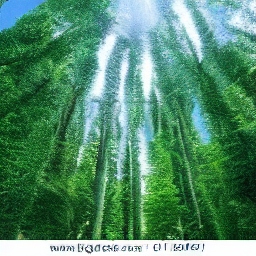}%
        &
        \scriptsize
        A forest with \textbf{vibrant green} trees all around.
    \end{tabularx}%
    \subcaption{}%
    \label{fig:study4:examples:c}
\end{subfigure}
\\[.25\baselineskip]
%
\begin{subfigure}[b]{0.32\textwidth}
\noindent
    \begin{tabularx}{\textwidth}{cX}%
      \includegraphics[width=\thumbnailwidth,valign=t]{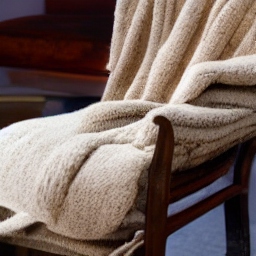}%
        \vspace{\skiplength}
        &
        \scriptsize
        A comfortable warm blanket resting on an antique rocking chair.
        \\
      \includegraphics[width=\thumbnailwidth,valign=t]{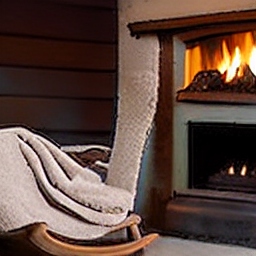}%
        &
        \scriptsize
        A comfortable warm blanket resting on an antique rocking chair \textbf{in front of a fireplace}.	
    \end{tabularx}%
    \subcaption{}%
    \label{fig:study4:examples:d}
\end{subfigure}
\begin{subfigure}[b]{0.32\textwidth}
    \begin{tabularx}{\textwidth}{cX}%
      \includegraphics[width=\thumbnailwidth,valign=t]{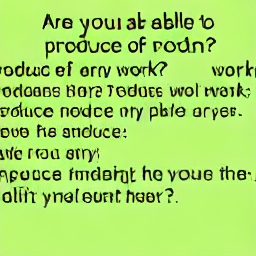}%
        \vspace{\skiplength}
        &
        \scriptsize
        Are you able to produce any of rodans work.	
        \\
      \includegraphics[width=\thumbnailwidth,valign=t]{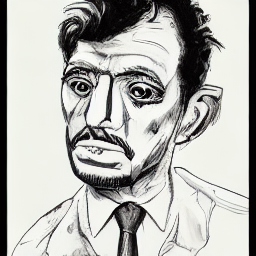}%
        &
        \scriptsize
        \textbf{Will you please correct} Rodans \textbf{art}work \textbf{for me}.
    \end{tabularx}%
    \subcaption{}%
    \label{fig:study4:examples:e}
\end{subfigure}
\begin{subfigure}[b]{0.32\textwidth}
    \begin{tabularx}{\textwidth}{cX}%
      \includegraphics[width=\thumbnailwidth,valign=t]{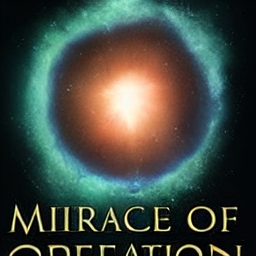}%
        \vspace{\skiplength}
        &
        \scriptsize
        miracle of creation	
        \\
      \includegraphics[width=\thumbnailwidth,valign=t]{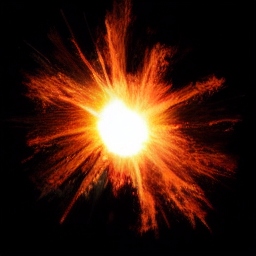}%
        &
        \scriptsize
        \textbf{explosion of creativity}	
    \end{tabularx}%
    \subcaption{}%
    \label{fig:study4:examples:f}
\end{subfigure}
\\[.25\baselineskip]
\begin{subfigure}[b]{0.32\textwidth}
\noindent
    \begin{tabularx}{\textwidth}{cX}%
      \includegraphics[width=\thumbnailwidth,valign=t]{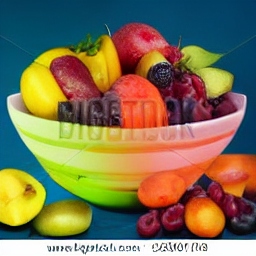}%
        \vspace{\skiplength}
        &
        \tiny
        A fruit bowl with vibrant colored fruits in it and a contrasting background	
        \\
      \includegraphics[width=\thumbnailwidth,valign=t]{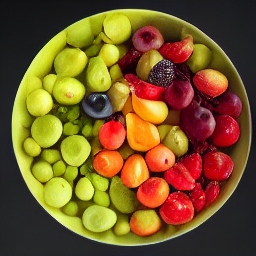}%
        &
        \tiny
        A \textbf{neutral colored} bowl with \textbf{a variety of several brightly} colored \textbf{and} vibrant fruits in it, and a background \textbf{that is darker to contrast with the fruit}.
    \end{tabularx}%
    \subcaption{}%
    \label{fig:study4:examples:g}
\end{subfigure}
\begin{subfigure}[b]{0.32\textwidth}
    \begin{tabularx}{\textwidth}{cX}%
      \includegraphics[width=\thumbnailwidth,valign=t]{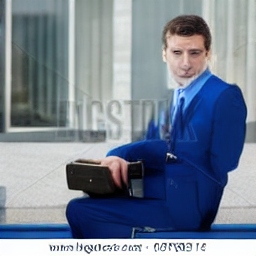}%
        \vspace{\skiplength}
        &
        \scriptsize
        A man in a blue business suit sitting on a bench. He holds a briefcase.
        \\
      \includegraphics[width=\thumbnailwidth,valign=t]{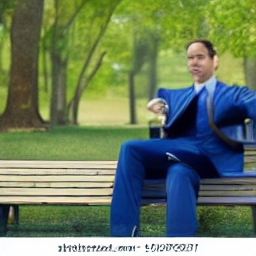}%
        &
        \scriptsize
        A man \textbf{seated on a park} bench. He \textbf{is in a} blue business suit. \textbf{A} briefcase \textbf{is beside him on the bench.}
    \end{tabularx}%
    \subcaption{}%
    \label{fig:study4:examples:h}
\end{subfigure}
\begin{subfigure}[b]{0.32\textwidth}
    \begin{tabularx}{\textwidth}{cX}%
      \includegraphics[width=\thumbnailwidth,valign=t]{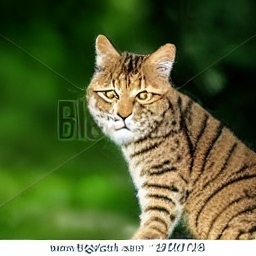}%
        \vspace{\skiplength}
        &
        \scriptsize
        A wild cat sitting on a brightly-painted fence.	
        \\
      \includegraphics[width=\thumbnailwidth,valign=t]{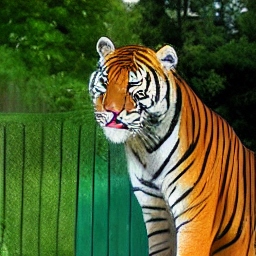}%
        &
        \scriptsize
        A \textbf{tiger stands on top of a} fence \textbf{that has been }painted \textbf{with vivid primary colors.}	
    \end{tabularx}%
    \subcaption{}%
    \label{fig:study4:examples:i}
\end{subfigure}
\caption{Examples of changes (highlighted in bold) in adjectives and adverbs (a--c), subjects (d--f) and multiple changes at once (g--i) made by crowd workers to their own prompts in Study~4.}
\label{fig:study4:images}%
\end{figure}%
}%

%

Our coding showed that the main strategy used by participants was modifying (i.e., adding, removing, or switching) adjectives in their prompts (see Figure~\ref{fig:study4:changes:cooccurence}).
For example, a participant changed the prompt \textit{``flowers in winter''} to \textit{``purple flowers in winter.''}
This was often combined with changes to the subject of the prompt (cf. Figure~\ref{fig:study4:images}), such as changing \textit{``sweeping arcs''} to \textit{``deep and broad, sweeping arcs in landscapes.''}
Some participants also adapted their prompts based on what they saw in the images, though this often resulted in only minor changes to the revised images.
    For instance, in the case of the above participant, the two images of mountainous landscapes were almost identical.
Another common approach was changing prepositions in the prompts. Few participants attempted to simplify their prompts, and relatively few made changes to cardinal numbers. For instance, one participant changed \textit{``draw a bunch of circles''} to \textit{``draw at least 15 circles,''} and another participant wanted to see \textit{``lots of puffy clouds''} without specifying the exact number.

We found that only one participant (the same as in Section \ref{sec:study3:modifiers}) demonstrated knowledge of prompt modifiers in all three of her prompts. An example written by this participants is
    \textit{``rainbow tyrannosaurus rex, \uwave{prehistoric landscape}, \uwave{studio ghibli}, \uwave{trending on artstation}}.''
This participant used the underlined prompt modifiers which are commonly used in the AI art community.
Only one other participant used a style modifier (\textit{``real photos of [...]''}) in one prompt.
This shows a very small increase in the use of prompt modifiers among participants in between Study 2 and Study 3, even though participants were specifically instructed to improve their artworks.

\subsubsection{Participants' revised images}%
We compared the two sets of images generated from each participant's prompts and found that over half of the revised sets showed no improvement in image quality (in terms of details, contrast, color, distortions, watermarks, and consistency).
Selected changes in the prompts and the resulting images are depicted in Figure~\ref{fig:study4:images}.
About half of the sets remained the same, 15\% were worse, and a third were better compared to the previous set.

Some participants were able to make improvements to the generated images, mainly by adding more details. Since participants added more tokens than they removed, the prompts were longer and resulted in about a third of the images having more details. Some participants also improved the images' colors and contrast by adding adjectives to the prompts. For instance, one participant improved the amount of details by adding \textit{``coral reef''} to the end of the prompt \textit{``scuba diver exploring unknown ocean.''} This change resulted in less blur and more details in the coral reef. However, strong changes in the style of the images were rare, with about 70\% of the revised sets being in the same or very similar style. Because participants did not use style modifiers, the revised images often resembled the initial images.

About 15\% of the images were of low aesthetic quality, often consisting of text with no discernible subject (see Figure~\ref{fig:study34-examples:b}).
These images were rarely improved between the studies, and when they were, it was often due to chance.
For instance, the subject was completely changed in about 10\% of the images. This was often a result of participants trying to have a conversation with the AI and entering a completely different prompt as input  (see Figure~\ref{fig:study4:examples:b} and Figure~\ref{fig:study4:examples:e}).

\subsubsection{Participants' use of negative terms}%
Nineteen participants used negative terms, with eight using them in all three prompts. In total, we collected 39~negative terms.
Many negative terms ($n=19$) aimed at removing or modifying the subject in various ways, such as removing \textit{``rocks''} from a beach, trying to correct a \textit{``weird face,''}  avoiding a \textit{``Nude, Naked, White, Man,''} removing the color \textit{``Green.''} in the image of a red star, or attempting to change the subject entirely (\textit{``ballroom''}).
Participants tried to change the style of the images in eight cases, using terms such as \textit{``Black, White, Colorless, Monochromatic,''} \textit{``opaque, solid,''} and \textit{``unfocused.''}
Four out of the 50~participants tried to remove text in the images, using negative terms such as \textit{``letters,''} \textit{``captions,''} and \textit{``text.''}
Only one participant attempted to remove watermarks, using the negative term \textit{``remove watermark.''}
As can be seen in this prompt, some participants did not understand the concept of negative terms, even though we explained it to them.
A few examples of failed and successful attempts are depicted in Figure~\ref{fig:study4:negativeterms}.
Some of the image generations failed, because the participant did not use the negative term correctly. For instance, the prompt on the bottom right of Figure~\ref{fig:study4:negativeterms} contains a monarch butterfly both in the prompt and negative term. The resulting image is sub-par compared to the image generated from the participant's original prompt.%
%
\newlength{\negwidth}%
\setlength{\negwidth}{1.6cm}%
\noindent%
\begin{figure}[bht]%
\centering%
\begin{tabularx}{\textwidth}{ccXccX}%
    \small before & \small after & \small revised prompt & \small before & \small after & \small revised prompt
\\
  \includegraphics[width=\negwidth,valign=t]{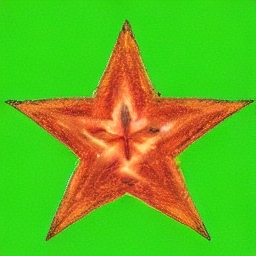}%
    &
  \includegraphics[width=\negwidth,valign=t]{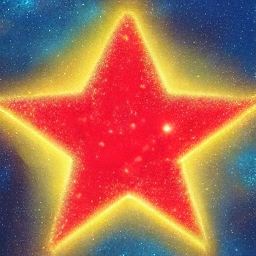}%
    &
    \footnotesize
    An explosively bright, dying star.
    \newline
    [Green]
&
  \includegraphics[width=\negwidth,valign=t]{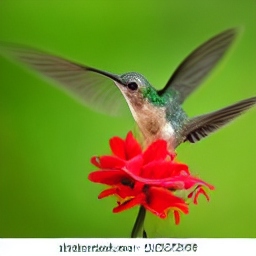}%
    &
  \includegraphics[width=\negwidth,valign=t]{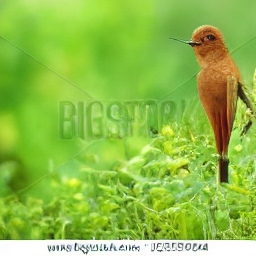}%
    &
    \footnotesize
    Hummingbird over red flower with a out of focus background of green shrubs	
    \newline
    [Hummingbird hovering over red flower with short focal point.]
    
\\[.25\baselineskip]

  \includegraphics[width=\negwidth,valign=t]{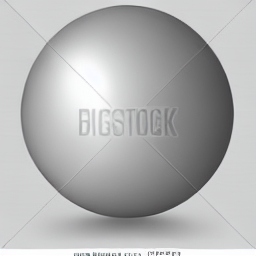}%
    &
  \includegraphics[width=\negwidth,valign=t]{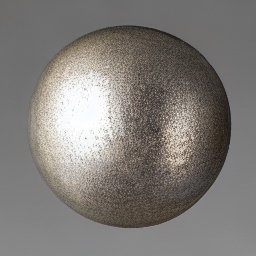}%
    &
    \footnotesize
    medium sized metallic sphere slightly above and left of the center of the image
    \newline
    [remove watermark]
&
  \includegraphics[width=\negwidth,valign=t]{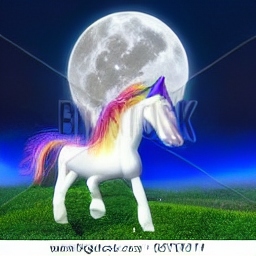}%
    &
  \includegraphics[width=\negwidth,valign=t]{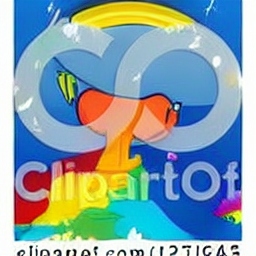}%
    &
    \footnotesize
    Rainbow colored unicorn with huge mane jumpin over the full moon	
    \newline
    [Shimmering hairy unicorn jumping over the full moon]
    
\\[.25\baselineskip]

  \includegraphics[width=\negwidth,valign=t]{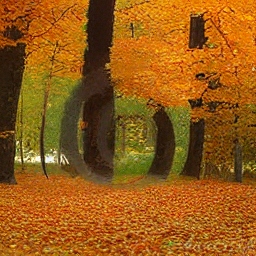}%
    &
  \includegraphics[width=\negwidth,valign=t]{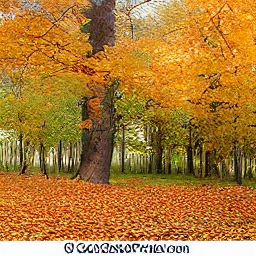}%
    &
    \footnotesize
    An autumn day in a colorful forest	
    \newline
    [Dark Circle]
&
  \includegraphics[width=\negwidth,valign=t]{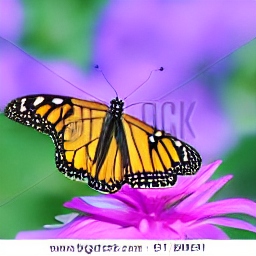}%
    &
  \includegraphics[width=\negwidth,valign=t]{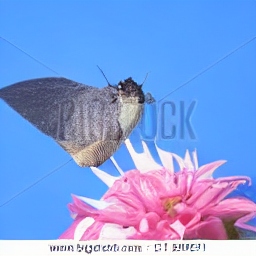}%
    &
    \footnotesize
    Butterfly Monarch on a flower	
    \newline
    [Monarch butterfly alit on a wildflower]
\\
\end{tabularx}%
\caption{Examples of successful (left) and failed (right) attempts of participants using negative terms (in square brackets).}
\label{fig:study4:negativeterms}%
\end{figure}%

\section{Discussion}%
\label{sec:discussion}%

In three studies, we explored the skill of prompt engineering with 227 participants recruited from a crowdsourcing platform.
Our first study shed light on whether laypeople have an understanding of what makes a ``good'' prompt.
%
Our findings 
indicate that participants can assess the quality of prompts and respective images. This ability increased with the participants' experience and interest in art.
In the subsequent two studies on writing and improving prompts, we found that participants wrote creative prompts in rich descriptive language which may result in beautiful digital artworks.
However, only a negligible amount of participants applied their knowledge of art in practice and failed to use terms commonly applied in communities of text-to-image art, such as Midjourney and Stable Diffusion. With the exception of one participant, these specific keywords were not (yet) part of the vocabulary of crowdsourced participants on MTurk at the time of our study. While the prompts written by participants were very descriptive and, in some cases, resulted in beautiful and interesting images, participants left the style of their image to chance. The prompts were missing modifiers that would more tightly control the style and quality of the image generation.
This applies to both the initial study on writing prompts and the study on revising and improving prompts.
In the latter, only a minority of the participants were able to improve their revised images, while most of the images remained about the same quality.
An overwhelming majority of participants left the style of the generated images to chance, even though they were specifically instructed to create ``artworks.''

Reflecting on these findings, it becomes evident that while crowd engagement with AI-driven art generation shows promise, there is a noticeable gap in the effective use of prompt engineering. This leads us to consider the broader implications and questions within the AI and Human-AI Interaction research fields. Particularly, it raises the question of whether prompt engineering is an intuitive skill that can be easily acquired or if it requires more specialized training and understanding.
%
%
\subsection{Prompt Engineering as a Non-Intuitive Skill}%
\label{sec:learnedskill}%
Our work adds to the discussion of broader research questions in the AI and Human-AI Interaction research communities:
    Can anyone become an artist with prompt engineering?
    Is prompt engineering a skill that is innate to us humans or is it a skill that needs to be acquired through practice and learning?
    If prompt engineering is an intuitive skill, how intuitive is it? Or in other words, how steep is the learning curve to prompt engineering?
These research questions have implications for the future of work and human-computer co-creativity~\citep{062-iccc20.pdf}.
If prompt engineering is an intuitive human skill that we humans can apply effortlessly, then we can look forward to a bright future where anyone can work in creative professions without having to develop special creative skills.
But if prompt engineering is a non-intuitive skill, its application could become limited to highly trained and skilled class of creative professionals who have mastered to speak the language of the generative model through extensive training.
The latter case could clearly negatively impact creative production and stifle innovation.

Prompting is a language-based practice and the use of language is intuitive to us humans. Therefore, one could assume that prompting is an intuitive skill.
It is easy to get started with writing prompts and prompting has a large potential in different fields and for many application domains.
However, our study found that effective prompt writing requires knowledge of keywords and key phrases. These prompt modifiers are an essential part of the skill of prompt engineering for AI generated art.
Typically, these keywords and key phrases are acquired through iterative experimentation and by learning from prompts shared in dedicated resources, on social media, or in online communities~\citep{aiartcreativity}.
Our studies empirically confirm that style modifiers are unknown to participants recruited on Amazon Mechanical Turk. Prompt modifiers that are being used profusely in the AI art community have not found their way into the collective vocabulary of crowdsourced participants on MTurk.
Participants in our study struggled to write and improve their prompts for the specific task of creating digital artworks.
This points towards prompt engineering being a non-intuitive skill or perhaps even a specialist skill.

However, we acknowledge this is a simplistic view.
One repeated interaction with the generative model does not lead to learning effects and a skill requires learning to be mastered. Perhaps a different question to ask is whether prompt engineering is a skill at all.
The initial outputs of text-to-image generation systems have a high randomness and, due to the ineffectiveness of discrete language for describing images in detail, it is very difficult to control the initial image. Therefore, text-to-image generation often requires several iterations to get to an acceptable level.
However, in practice, this first image is just the starting point. There is much more to prompt engineering than just writing textual prompts. In particular, the first image is generally only the start in a longer creative process that can include, for instance, ControlNet, image editors, image-to-image generation, inpainting and outpainting, facial detailing, and upscaling. Upscaling in itself is often completed with generative models, such as Real-ESRGAN \citep{RealESRGAN}, thereby introducing artifacts that may not be wanted, requiring further editing.
Prompt engineering is not just tied to the use of a platform, but to a whole ecosystem of creative tools (e.g., image editors and generative models).
Learning this variety of tools requires specific training.
%
We speculate on four possible futures for the skill of prompt engineering in the following section.%

%
%
\subsection{On the Future of 
Creative Production with Prompt Engineering}%
\label{sec:creativeeconomy}%
%
Text-to-image generation opens new opportunities for creative production of digital images and artworks.
Whether prompt engineering will become an expert skill or even a novel profession is still open.
In this section, we speculate on four possible futures of prompt engineering.
%
\subsubsection{Prompt engineering as an expert skill}%
In the future, prompt engineering could become an expert skill that requires deep subject-matter expertise (e.g., knowledge of subject-specific keywords, prompt modifiers, and their combinations, but also of the idiosyncrasies of the training data and system configuration parameters) to effectively control the output of generative systems. This is similar to the move in the field of machine learning towards ``foundation models'' \citep{FoundationModels}.
Foundation models are very large and costly to train, operate, and maintain. As a result, the creation of these models as well as research on these models is limited to a small number of well-financed research institutes that employ highly-skilled professionals working in well-funded research institutes and organizations. If prompt engineering becomes a highly skilled profession, it may become exclusive to a narrow group of privileged individuals who have undergone extensive training.

In our study (conducted in mid-2022), only one participant demonstrated knowledge of prompt modifiers.
Controlling text-to-image generation is still a difficult task, and practitioners spend many hours obsessing over very small details. In fact, as described in the previous section, the workflow of text-to-image generation typically involves more steps than just prompt writing -- this can be considered only the first steps in a complex task work flow that involves image generation models as well as editors and specialized tools.
From this perspective, text-to-image generation remains an expert skill that, while now accessible to a large part of the population, remains difficult to master. This perspective is supported by the finding that most images generated in this study failed to achieve the given goal of creating an ``artwork'' (without defining what an artwork is and, granted, acknowledging that photographs are also art).
However, prompt modifiers that would nudge the text-to-image generation system to produce artworks, whether photography or oil painting, were notably absent in the prompts written by the participants in our study.
%
\subsubsection{Prompt engineering as an everyday skill}%
In the future, prompt engineering could become a common practice.
In this scenario, people would adapt their creative practices and language to facilitate effective interaction with AI because it is a skill that is needed in everyday life.
People have a need for visual content, and AI-generated content could satisfy this need, from internet memes to the design of greeting cards, logos, and artworks. Prompt engineering could also be used for self-actualization, creativity, and therapy to improve mental health and well-being ~\citep{BURLESON2005436}.
In this scenario, people would expand their existing vocabulary to include terms used in prompt engineering in order to produce meaningful outcomes with generative systems. Learning prompt engineering would be similar to learning a new language. This skill would be acquired at an early age, just like internet literacy is today acquired effortlessly by the generation born after the internet found widespread use.
The skill could also become part of media literacy education in schools to elevate the common skill-level to a professional level.

There is some support in our studies for this speculative future.
Most participants were able to write creative and detailed prompt.
Digital literacy in using text-to-image generation systems can therefore be assumed to be present to some degree. With text-to-image technology improving, the skill of prompt engineering may become easier and not require as many special keywords and modifiers in the prompts.
On the other hand, if this skill was to be used daily, participants would likely want to improve their skill to make use of prompt engineering more efficiently.
Our study itself is an example of how rapidly the world changed after the wide-spread release of generative AI. It is likely that our study would have a different outcome today, simply because more people, today, know how to write basic prompts and perform basic prompt engineering.
This would support the perspective of prompt engineering becoming a basic everyday skill.

\subsubsection{Prompt engineering as an obsolete skill}%
In the future, prompt engineering could become irrelevant.
How users interact with AI models is closely coupled with the advances of AI technology and what the AI models are capable of.
Prompt engineering can be seen as the smell of a half-baked product that does not solve its users' needs.
As generative systems improve their ability to understand the intent of users, prompt engineering could be a short-lived trend.
The problem of aligning AI with human intent is known as AI~alignment in the scholarly literature \citep{Gabriel2020_Article_ArtificialIntelligenceValuesAn.pdf}.
State-of-the-art systems, such as ChatGPT~\citep{chatgpt} and 
DALL-E~3 \citep{dalle-3}, demonstrate impressive performance in understanding textual input prompts and aligning with user intent. With these systems, users of all skill levels can generate content from textual prompts.
As generative systems become better at understanding user intent, prompt engineering could become obsolete, similar to how we no longer use block printing.
Prompt engineering could become unnecessary --- an archaic skill that does not require expert training and that only few people exercise for nostalgic reasons.

Support for this speculative future comes primarily from the creators of generative tools themselves.
    One example is Midjourney which introduced a noticeable jump in quality between version 5 and 6, with the latter version requiring fewer keyword modifiers in prompts \citep{mjfewerkeywords}.
    The CEO of Midjourney, David Holz, has recommended not to put too much effort into acquiring the skill of prompt engineering because generative systems change too quickly \citep{holz_office_hours,pefuture}.
    Another example of the skill potentially becoming obsolete are language models that generate prompts. This approach is used in some text-to-image generation systems to rewrite the user input and improve the prompt quality. One instance of such LM-based prompt rewriting is used on the image generator ideogram.ai.
    Such prompt rewriting could, if language models were well aligned with user intent, make the skill of prompt engineering obsolete.

\subsubsection{Prompt engineering as personal signature or curation skill}%
In the future, our human senses could become better at distinguishing hand-crafted art from AI-generated digital art.
AI artist Mario Klingemann speculated that with the influx of AI-generated images, this skill would help us notice subtle nuances, details, and imperfections in AI-generated art, which could become more important in determining the aesthetic quality of an art piece.\footnote{https://twitter.com/quasimondo/status/1512769106717593610}
In this scenario, anyone could write prompts for generative systems with good results, but only a few would become masters of prompt engineering. The practice of prompt engineering would remain a necessary skill for applying finishing touches and optimizing generative results, as well as imbuing an artwork with a personal style to distinguish it from bland ``off-the-shelf'' generations. Alternatively or in parallel, prompt engineering could evolve into a curation skill~--- a personal practice in which everyone has their own curated sets of textual and visual inputs used to fine-tune generative models for different purposes.
In this scenario, the machine would personalize and adapt to humans, rather than the other way around.
As a result, the generative machine would potentially develop a perfect understanding of user intent.
Current approaches that cater towards this future are
Low-Rank Adaptation (LoRA) \citep{hu2022lora},
Dreambooth \citep{dreambooth} and textual inversion \citep{textualinversion}.

There is some support for this possible future in our study.
Participants were able to tell bad quality prompts from good quality prompts. This could enable participants to curate prompts.
Furthermore, the prompts from some of our study participants had a clear personal signature. The idiosyncratic way of writing prompts was easily recognizable in some participants.
Future generative models could pick up on these subtleties and adapt to the idiosyncrasies of their user-written prompts. This would, clearly, enable these models to outperform models that cannot adapt to the subtleties of user intent.

\subsubsection{Review and outlook}
\label{sec:skillfuture:summary}
These four speculative futures were initially formulated in 2022. Two years on, there has still not been a definitive answer to the question which potential future is more likely and in which direction we are heading.
School curricula have still not adapted prompt engineering.  Prompt engineering is a perishable skill with a short shelf life. Every time a model is updated, practitioners have to adopt new tricks and the skill of prompt engineering has to be learned anew. Even the CEOs of generative AI companies, such as Midjourney, recommend not putting too much effort into learning the skill of prompt engineering, because it is changing too rapidly \citep{holz_office_hours,mjfewerkeywords}.
On the other hand, what is enduring is the continued public interest in prompt engineering and its usefulness for solving real-world problems, such as question--answering over text documents \citep{RAG,oppenlaender2023mapping}.
   LM-based agents have become popular in industry due to their usefulness in solving difficult problems, surpassing even web development frameworks in popularity \citep{jufo3,wu2024autogen}. From this perspective, prompt engineering -- and, more generally, AI engineering -- is considered a valuable skill to have and teach in school curricula \citep{AISKILLINDUSTRY}.
 There are now job positions made available with title `Prompt Engineer,' not only at technology companies, but also in the public sector, such as the position of a `Senior Prompt Engineer' role at the AI Safety Institute of the UK's Department for Science, Innovation \& Technology \citep{promptengineer}.
However, digging deeper into this particular job role, it is clear that the role of Prompt Engineer is more encompassing than just writing prompts -- it requires technology skills, such as Python programming, and knowledge of deep learning and evaluation frameworks. Advanced prompting techniques are just one aspect in this job profile. 

The rapidly evolving landscape of generative AI is the crux of the skill of prompt engineering.
While the skill is useful for hobbyists and practitioners in industry, it is changing too rapidly to make it a sensible addition to school curricula.
As a consequence, we can expect the gap between academia and practice to become even wider in the coming years.
%
With this context in mind, we now turn our attention to future~work.
\subsection{Future Work}%
\label{sec:futurework}%

We conducted an experiment asking crowdsourced participants to write, improve, and assess prompts.
The study was conducted in May--July 2022, at a time when text-to-image generation was still unknown to most people.
Today, many people have tried text-to-image generation at least once, and many more have heard of the scandals and lawsuits surrounding generative AI.
Knowledge of these scandals may shape the perception and use of generative AI \citep{useofAI}.
Our study was conducted in mid-2022, and, today, it would be difficult to recruit a participant sample as naive to text-to-image generation as our sample.
This makes our study important and valuable.
This is precisely the value of our work: we evaluated prompt engineering as a skill with naive participants,
at a time when text-to-image generation was not as popular as it is today.
    Midjourney was just released and still unknown to many people. Stable Diffusion was released months later, in August 2022. And ChatGPT was made available only later the same year.
Our participant sample consisted of laypeople who were still naive to ``prompt engineering'' for text-to-image generation.
Such a participant sample is hard to find today -- most people have heard and tried image generators or language models at least once, and ChatGPT is available for free.
The interaction with image generators 
is typically interactive: a user enters a prompt, observes the results, and types another prompt in response to the observed output.
Therefore, text-to-image generation has learning built in. It is, today, very difficult to disentangle this aspect of interactive learning from any investigation concerned with understanding the skill underlying prompt engineering.
In our study, participants had no experience in text-to-image generation (except for one participant) and underwent one repeated interaction with the system. This allowed us to investigate the skill of prompt engineering in our experiment.
Future work should build on this, for the reasons outlined in Section \ref{sec:skillfuture:summary}, even though recruiting naive participants is challenging.
Is is pertinent to investigate whether prompt engineering is to be included in school and high education curricula. To make an informed decision, one would need to know more about how quickly the skill of prompt engineering is learned, what it exactly entails (besides prompt writing), the steepness of the learning curve, and how long it would take to master this skill. All these factors are needed to make an informed decision of whether prompt engineering is a skill to include in curricula.

We investigated textual prompt engineering in our paper. But there are more facets to text-to-image generation in practice, as discussed in Section \ref{sec:learnedskill} and outlined in \citet{aiartcreativity}, such as visual prompting and configuration parameters in text-to-image generation systems.
Future research should consider the complexity and diversity of skills and should always consider all facets involved in the creative process, not just textual prompts.
%
Future work could envision how the machine can adapt to humans and possibly provide guidelines for AI researchers, so they can develop AI models that understand user intent and foster a more ``intimate'' relationship with users \citep{p75-weiser.pdf}.
Language -- whether written or spoken -- is the perfect vehicle for this intimate relationship with users.

\subsection{Limitations}
\label{sec:limitations}%
We acknowledge a number of risks to the validity of our exploratory studies.
Aesthetic quality assessment, as explored in Study 1, inherently carries a high degree of subjectivity \citep{joshi_1.pdf}. Various factors influence such ratings, including personal values, background, image content and its interestingness, contrast, proportion, number of elements, novelty, and appropriateness \citep{joshi_1.pdf,1606.01621.pdf,DS77_158.pdf}.
We acknowledge that the task given to participants in Study~1 was challenging, particularly in terms of visualizing and evaluating the potential outcome of a written prompt. While our findings indicate that participants could discern between low and high-quality prompts, we must also recognize a key limitation: the absence of direct measurements for the actual quality of the generated artworks themselves. This gap points to the difficulty in quantitatively assessing artistic creations, a challenge further compounded by the subjective nature of aesthetic evaluation. Future research could benefit from developing more concrete and objective criteria or methods to assess the quality of prompt--artwork pairs, providing a more comprehensive understanding of the relationship between prompt quality and the resulting art.

We further acknowledge limitations in our choice of text-to-image generation system.
Our 
main motivation for selecting Latent Diffusion was, at the time, that it was a state-of-the-art image generation system allowing the deterministic and reproducible generation of images. We tested a dozen of other text-to-image generation systems, notably CLIP-Guided Diffusion, GLID-3-XL, DALL-E mini (Craiyon), Latent Majesty Diffusion, DISCO Diffusion, and VQGAN-CLIP \citep{VQGAN-CLIP}, with mixed results.
Some systems had non-deterministic (random) components leading to image generations not being reproducible. Other systems were not advanced enough, at the time, to produce recognizable results.
Note, however, that while Latent Diffusion is a powerful image generation system, it may respond differently to style keywords than CLIP-guided systems.
Our choice of latent diffusion was a compromise between reproducibility and performance.
In any case, we argue the choice of image generation system does not matter much for our study. Our study setup did not provide interactive feedback to participants, and the system itself was not our subject of study. Instead, we were interested in the intrinsic skill of participants.
However, only one participant used specific keywords (prompt modifiers) in our study. The second round of images was also never shown to participants. Therefore, we can safely assert that the choice of system had no effect on how participants adapted to the generative system and how they wrote and revised their prompts.%
%

Regarding the observation about participants making greater improvements in prompts but not in modifiers in Study~3, we realize that our initial instructions did not explicitly guide participants to modify the stylistic elements or modifiers of the prompts. This oversight might have led to less emphasis on altering these aspects, thus skewing the results in favor of more noticeable changes in the prompts' substantive content.
However, the overall tasks given to the participant was to improve the artwork, and the instructions were thus implicitly part of the given task.
As an alternative approach, a more directed instruction could have provided insights into how novice users interact with and perceive the importance of prompt modifiers in text-to-image generation. This realization could be a valuable consideration for the design of future studies.

Last, we acknowledge that to explore the dynamics of prompting skill learning, a long-term field study would need to be conducted. However, our one time modification experiment provides a first indication indicating that the skill of prompting is not an innate skill that users can apply without learning about it first.
Recent work by \citet{donyehiya2023human} provides intriguing insights on the dynamics of prompt learning on Midjourney that future work could build on.%

\section{Conclusion}%
\label{sec:conclusion}%
The past few years have seen the rise of generative models.
It is too early to tell whether this development will give birth to new professions, such as ``prompt engineer.''
However, generative AI will deeply affect and reconfigure the fabric of our society.
This opens exciting opportunities for research in the field of HCI.

This article investigated prompt engineering, as a new type of skill,  in the context of AI art.
In three studies, we investigated whether laypeople naive to text-to-image generation
could recognize the quality of prompts and their resulting images, and whether participants could write and improve prompts, without repeated feedback from the image generation system.
We found participants recruited from
a crowdsourcing platform
were creative and able to write prompts for text-to-image generation systems in rich descriptive language, but lacked the special vocabulary found in AI art communities. The use of prompt modifiers was not intuitive to participants, pointing towards prompt engineering being a non-intuitive skill.
We discussed the importance of our study's findings, and speculated on four possible futures for prompt engineering.
We hope that whatever the landscape of creative production will turn out to be in the future, it will be an inclusive creative economy in which everyone 
can participate in meaningful ways.


\bibliographystyle{apacite}%
\bibliography{manuscript}%

\appendix
\newpage
\section{Set of images used in Study 1}
\label{appendix:images}

\newlength{\imageheight}
\setlength{\imageheight}{4.5cm}
\newlength{\skipspace}
\setlength{\skipspace}{.1cm}

\subsection{Images with High Aesthetic Appeal}
\label{appendix:a1}

\noindent
\begin{tabularx}{\textwidth}{
    >{\hsize=.245\hsize}X
    >{\hsize=.245\hsize}X
    >{\hsize=.245\hsize}X
    >{\hsize=.245\hsize}X
}
      \vspace{0pt} \includegraphics[width=.24\textwidth,valign=t]{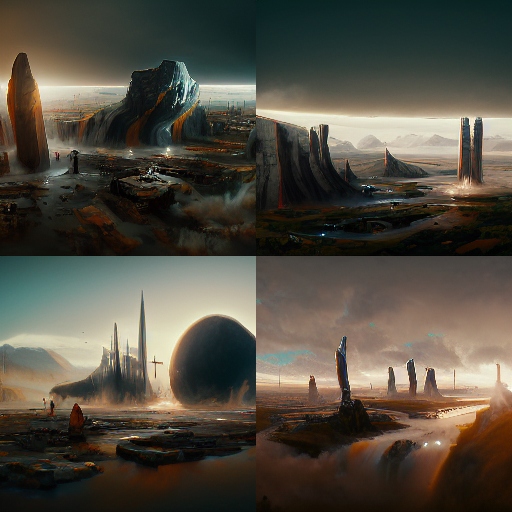}
    &
      \vspace{0pt} \includegraphics[width=.24\textwidth,valign=t]{figures/midjourney/255acf24-263a-4625-93c1-b5757b30e702x.jpg}
    &
      \vspace{0pt} \includegraphics[width=.24\textwidth,valign=t]{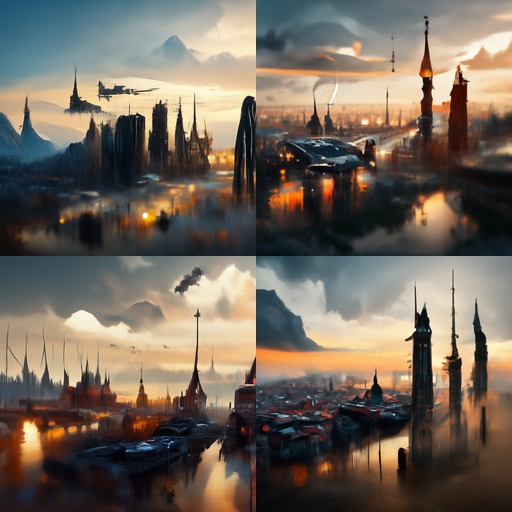}
    &
      \vspace{0pt} \includegraphics[width=.24\textwidth,valign=t]{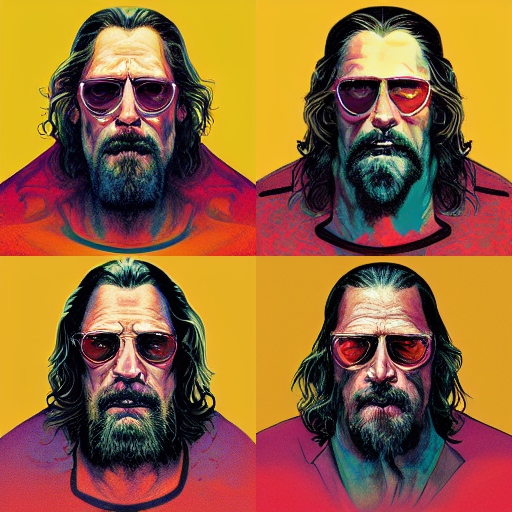}
\\
      \footnotesize
      \textbf{H1:} \prompt{25a8c8c9-55c2-48ac-af30-3e1333eb6b76x.jpg}
    &
      \footnotesize
      \textbf{H4:} \prompt{255acf24-263a-4625-93c1-b5757b30e702x.jpg}
    &
      \footnotesize
      \textbf{H5:} \prompt{0f0c96fc-2160-4e59-8331-1bd1bd619ad5x.jpg}
    &
      \footnotesize
      \textbf{H6:} \prompt{2be06581-4f12-4b8b-87a4-f77b9e4b375cx.jpg}
\\
\end{tabularx}

\noindent
\begin{tabularx}{\textwidth}{
    >{\hsize=.33\hsize}X
    >{\hsize=.33\hsize}X
    >{\hsize=.33\hsize}X
}
      \vspace{0pt} \includegraphics[width=.325\textwidth,valign=t]{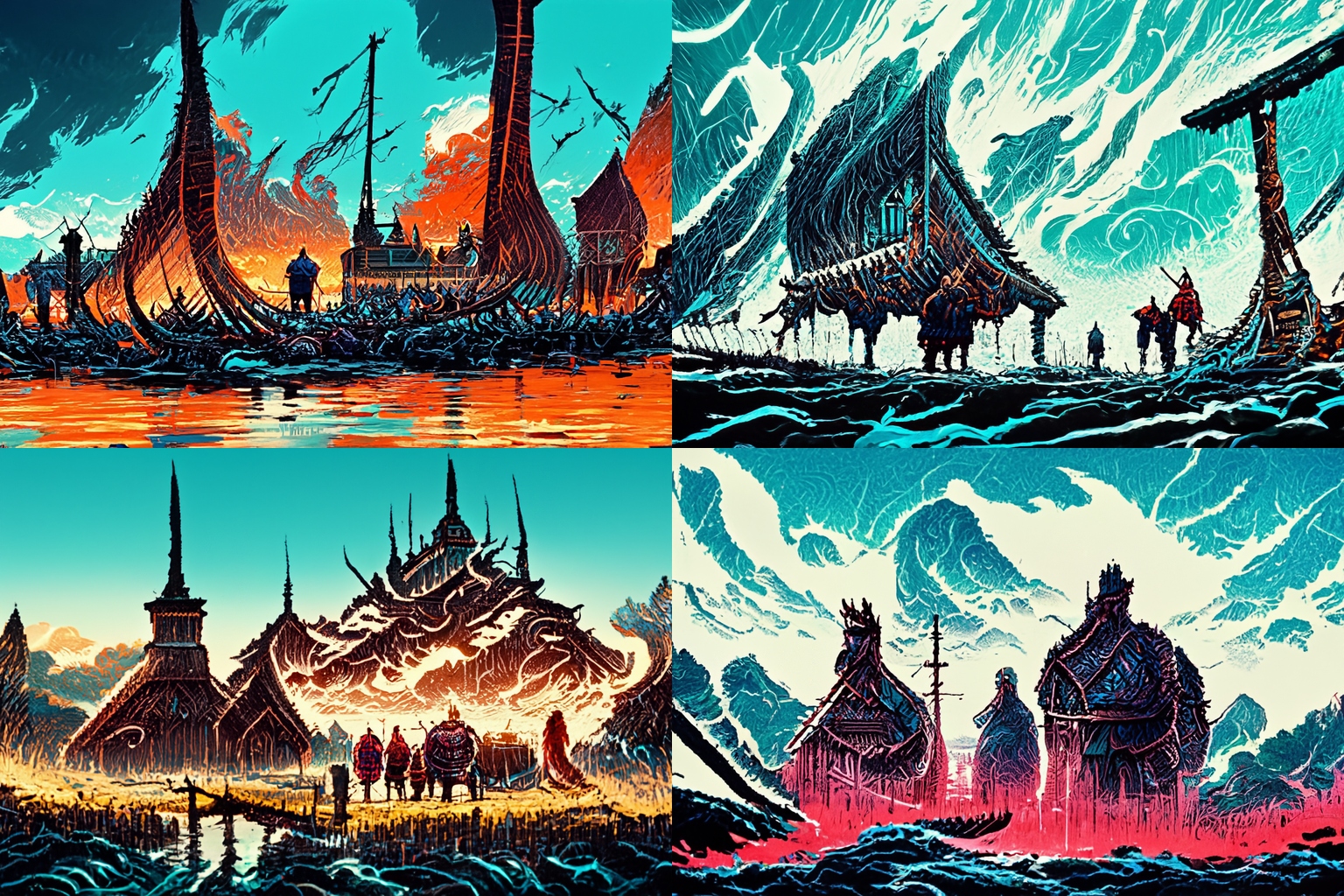}
    &
      \vspace{0pt} \includegraphics[width=.325\textwidth,valign=t]{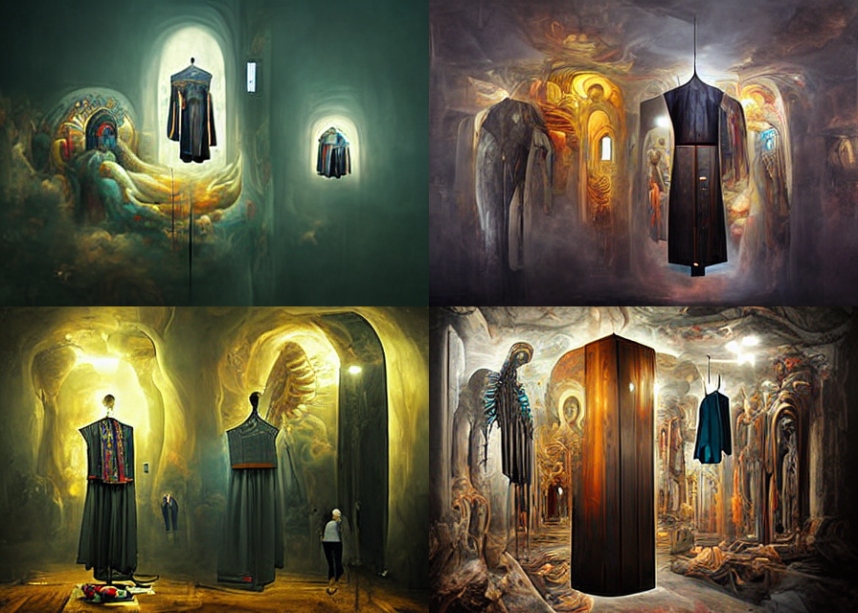}
    &
      \vspace{0pt} \includegraphics[width=.325\textwidth,valign=t]{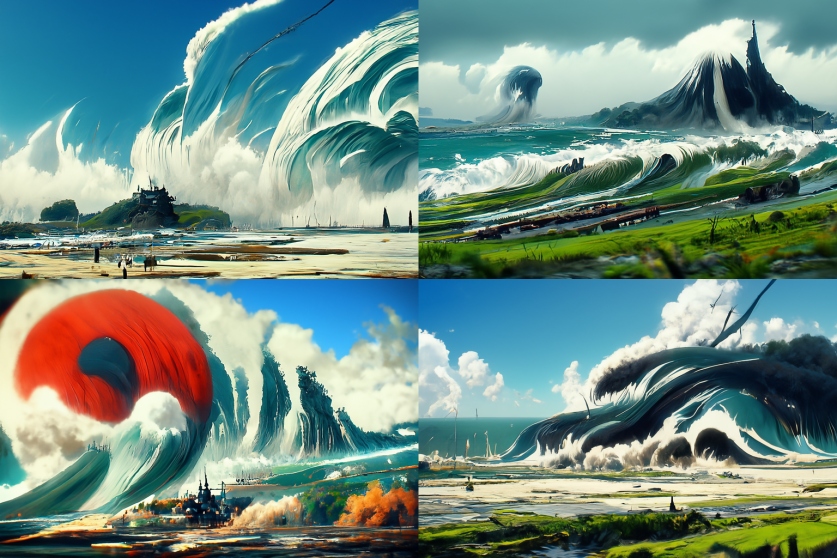}
\\
      \footnotesize
      \textbf{H2:} \prompt{a7cbe696-4af4-4af6-9224-c4447c29831cx.jpg}
    &
      \footnotesize
      \textbf{H7:} \prompt{7ca2d0bd-6df7-49c1-91c3-916726216dd8x.jpg}
    &
      \footnotesize
      \textbf{H9:} \prompt{896c8280-187b-4b55-a640-9157a9af9a6dx.jpg}
\\
\end{tabularx}

\noindent
\begin{tabularx}{\textwidth}{
    >{\hsize=.245\hsize}X
    >{\hsize=.245\hsize}X
    >{\hsize=.49\hsize}X
}
      \vspace{0pt} \includegraphics[width=.24\textwidth,valign=t]{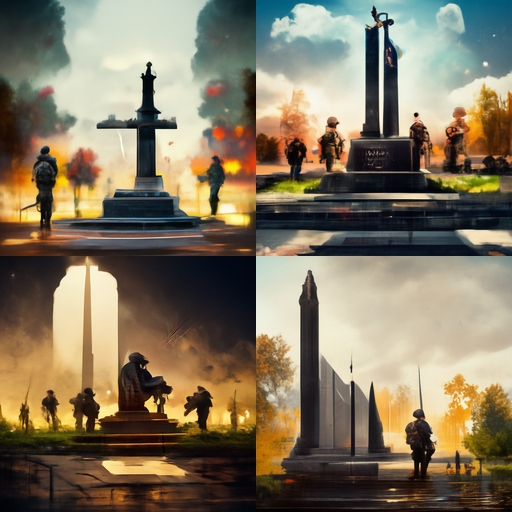}
    &
      \vspace{0pt} \includegraphics[width=.24\textwidth,valign=t]{figures/midjourney/22a357e6-2425-42b7-8475-dd05c419ecc9x.jpg}
    &
      \vspace{0pt} \includegraphics[width=.485\textwidth,valign=t]{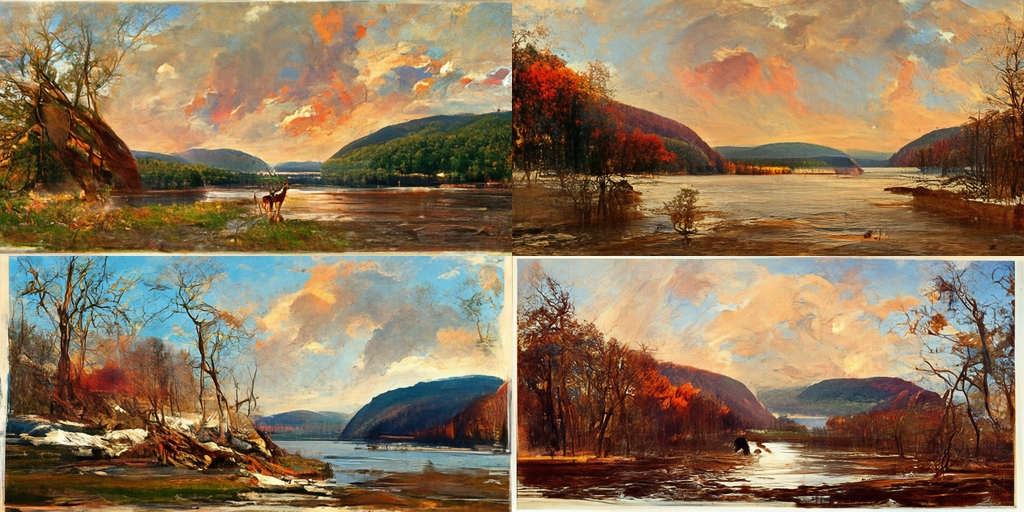}
\\
      \footnotesize
      \textbf{H8:} \prompt{5f98723a-488e-493e-954b-9cbf38baa69dx.jpg}
      &
      \footnotesize
      \textbf{H10:} \prompt{22a357e6-2425-42b7-8475-dd05c419ecc9x.jpg}
      &
      \footnotesize
      \textbf{H3:} \prompt{fb2621f1-b3f0-4cd5-8f93-774836ff1ecex.jpg}    
\end{tabularx}



\newpage
\subsection{Images with Low Aesthetic Appeal}
\label{appendix:a2}

\noindent
\begin{tabularx}{\textwidth}{
    >{\hsize=.245\hsize}X
    >{\hsize=.245\hsize}X
    >{\hsize=.245\hsize}X
    >{\hsize=.245\hsize}X
}
      \vspace{0pt} \includegraphics[width=.24\textwidth,valign=t]{figures/midjourney/66b13a9f-9c0c-4c9f-aa31-d9c0f2aa8324x.jpg}
    & 
      \vspace{0pt} \includegraphics[width=.24\textwidth,valign=t]{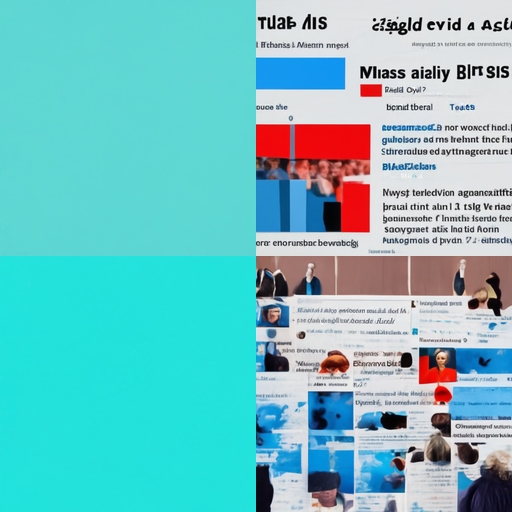}
    & 
      \vspace{0pt} \includegraphics[width=.24\textwidth,valign=t]{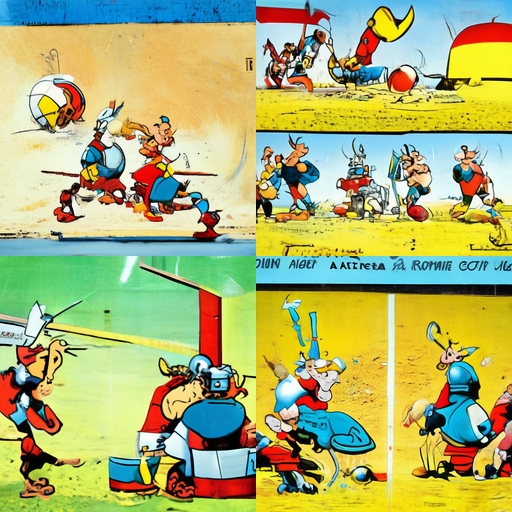}
    &
      \vspace{0pt} \includegraphics[width=.24\textwidth,valign=t]{figures/midjourney/8d91ae8d-673b-40e0-abed-4ce965353c16x.jpg}
\\
        \footnotesize
        \textbf{L1:}
        \prompt{66b13a9f-9c0c-4c9f-aa31-d9c0f2aa8324x.jpg}
    & 
        \footnotesize
        \textbf{L2:}
        \prompt{6c0066ee-61d7-44a1-bef8-24908db1af47x.jpg}
    & 
        \footnotesize
        \textbf{L3:}
        \prompt{f41ba10e-05f8-458a-8709-22d1060ea5b2x.jpg}
    &
        \footnotesize
        \textbf{L4:} 
        \prompt{8d91ae8d-673b-40e0-abed-4ce965353c16x.jpg}
\\
\end{tabularx}

\noindent
\begin{tabularx}{\textwidth}{
    >{\hsize=.245\hsize}X
    >{\hsize=.245\hsize}X
    >{\hsize=.245\hsize}X
    >{\hsize=.245\hsize}X
}
      \vspace{0pt} \includegraphics[width=.24\textwidth,valign=t]{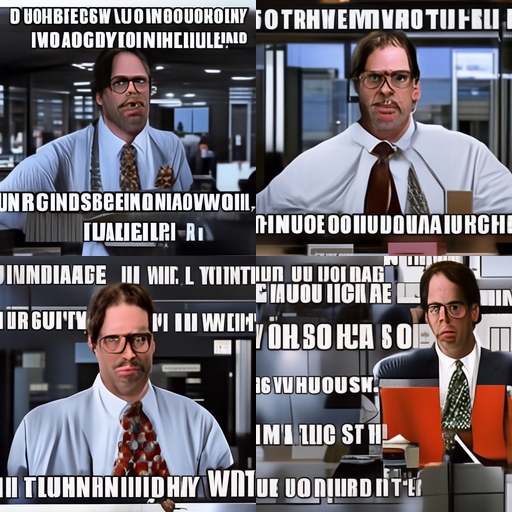}
    &
      \vspace{0pt} \includegraphics[width=.24\textwidth,valign=t]{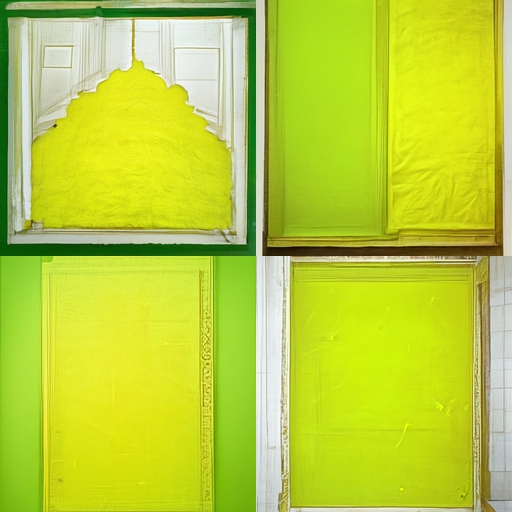}
    &
      \vspace{0pt} \includegraphics[width=.24\textwidth,valign=t]{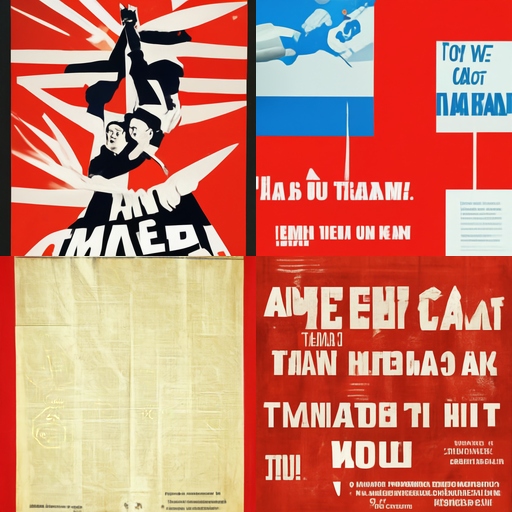}
    & 
      \vspace{0pt} \includegraphics[width=.24\textwidth,valign=t]{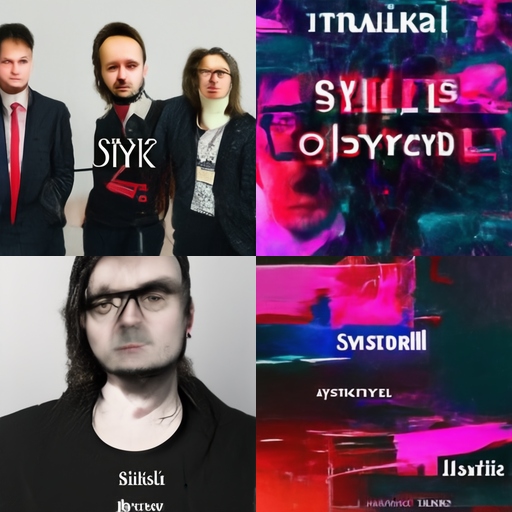}
\\
        \footnotesize
        \textbf{L5:} 
        \prompt{71dfc24b-d290-4500-8047-7a0da7bd917bx.jpg}
    &
        \footnotesize
        \textbf{L6:}
        \prompt{1c67fdcd-b16b-44ba-8f82-16f5f443f7c9x.jpg}
    &
        \footnotesize
        \textbf{L7:}
        \prompt{83328d03-585a-4351-b48f-0bfa1cbe72c0x.jpg}
    &
        \footnotesize
        \textbf{L8:}
        \prompt{6d8ace44-d363-4654-8e95-0f266bf5886dx.jpg}
\\
\end{tabularx}

\noindent
\begin{tabularx}{\textwidth}{
    >{\hsize=.245\hsize}X
    >{\hsize=.245\hsize}X
    >{\hsize=.245\hsize}X
    >{\hsize=.245\hsize}X
}
      \vspace{0pt} \includegraphics[width=.24\textwidth,valign=t]{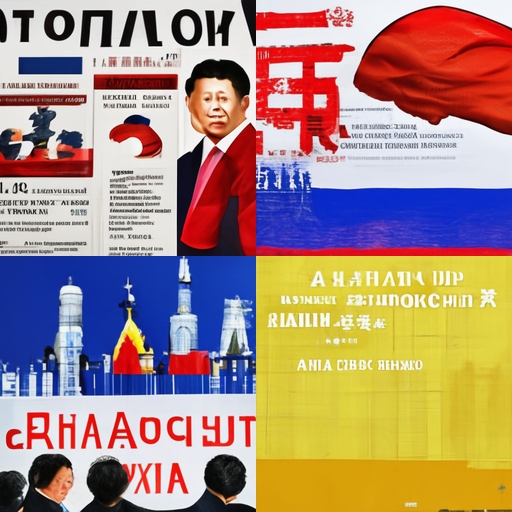}
    &
      \vspace{0pt} \includegraphics[width=.24\textwidth,valign=t]{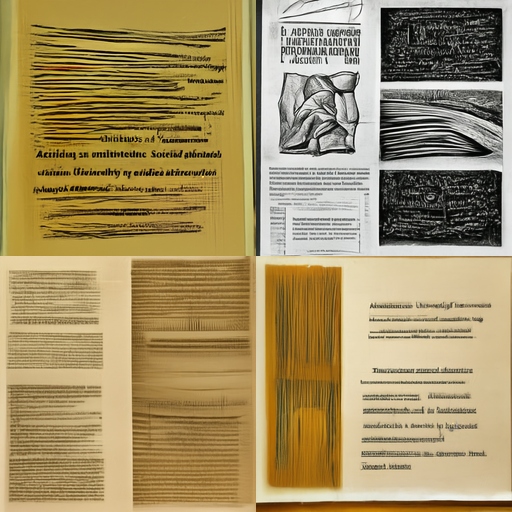}
    &
    &
\\
        \footnotesize
        \textbf{L9:} 
        \prompt{46ae15d3-20b6-4cf5-914e-db537e760788x.jpg}
    &
        \footnotesize
        \textbf{L10:}
        \prompt{0da6b738-eb79-4b41-b212-2c3a97ada099x.jpg}
    &
    &
\end{tabularx}

\end{document}